\documentclass[nofootinbib,showpacs,aps,prl,reprint,letterpaper]{revtex4-2}
\usepackage{amsmath,amssymb}
\usepackage{epsfig}
\usepackage{graphicx}
\usepackage{amsmath}
\usepackage{slashed}
\usepackage{amsfonts}
\usepackage{epstopdf}
\usepackage{tablefootnote}

\usepackage{graphicx,subfigure}
\usepackage{amsmath,amssymb}
\usepackage{hhline}

\usepackage[pdftex, pdfborder= 0 0 0, citecolor=blue, urlcolor=blue, linkcolor=blue, colorlinks=true, bookmarksopen=true]{hyperref}       
\newcommand{\ssec}[1]{\emph{#1}.---}                               %

\begin{document}
	
	\title{Nuclear state and level densities of actinides in the shell-model Monte Carlo}
	
	\author{D. DeMartini and Y. Alhassid}
	
	\affiliation{Center for Theoretical Physics, Sloane Physics Laboratory, Yale University, New Haven, Connecticut 06520, USA}

	\begin{abstract}
	
	Actinides are of great interest in astrophysics and technology applications since they can fission. However, the microscopic calculation of their statistical properties in the presence of correlations poses a major theoretical challenge. The configuration-interaction shell-model is a suitable framework to calculate these properties but the required large model spaces are beyond the reach of conventional diagonalization methods. The shell-model Monte Carlo (SMMC) method enables calculations in very large model spaces and was applied to nuclei as heavy as the lanthanides. Here, we extend the SMMC method to the actinides. Fifteen even-even and odd-mass actinides $^{232}$Th, $^{\textrm{234-239}}$U, $^{\textrm{240-243}}$Pu, $^{\textrm{246-248}}$Cm, and $^{250}$Cf are studied using a single-particle model space that is larger than one major shell each for protons and neutrons, with a total dimension of the many-particle space as large as $10^{32}$.  We calculate nuclear state densities of these actinides and find they are strongly enhanced in comparison with mean-field densities.  We use spin projection methods to calculate nuclear level densities and average $s$-wave neutron resonance spacings, both of which are found to be in good agreement with experiments. 
	\end{abstract}
	\maketitle
	
\ssec{Introduction}	The nuclear level density (NLD) is a fundamental statistical property of nuclei with importance to compound-nucleus reactions~\cite{Hilaire2021}, astrophysical abundances and processes~\cite{Arnould2007}, and more. Theoretical calculations of the statistical properties of heavy open-shell nuclei have required either highly truncated model spaces, or the use of mean-field approximations (e.g., density function theory~\cite{Drut2010}) which miss important correlations between nucleons. Accurate calculations of statistical properties in the actinide mass region ($A \sim 240$) require configuration-interaction (CI) shell model spaces larger than $10^{30}$, while computational costs currently limit conventional diagonalization methods to model spaces around $10^{11}$~\cite{Tsunoda2017,Shimizu2019}. The shell-model Monte Carlo (SMMC)~\cite{Alhassid2017} method enables microscopic calculations of nuclear properties in the presence of correlations in shell model spaces that are many orders of magnitude larger than what can be treated in conventional methods. Previously, the SMMC method has been applied to nuclei as heavy as the lanthanides with $A \sim 160$~\cite{Alhassid2008,Ozen2013,Ozen2015,Alhassid2014,Guttormsen2021,Mercenne2024,DeMartini2025}. 
	
	Several experimental methods have been used to measure NLDs. The Oslo method~\cite{Schiller2000,Guttormsen2005,Larsen2011} has been used to measure the NLDs of nuclei across a broad region of the nuclear chart at excitation energies below their neutron separation energies $S_n$. In particular, the Oslo method has been used to measure NLDs in several actinides~\cite{Guttormsen2013,Tornyi2014,Laplace2016,Zeiser2019}. 
	
	Actinides can fission and the deformation dependence of their NLDs is an important input to models of nuclear fission~\cite{Mazurek2011,Ward2017}. Additionally, the quadrupole deformation of $^{238}$U, the largest nucleus collided at the Relativistic Heavy Ion Collider, can impact observables like the transverse momentum distributions of particles detected in relativistic heavy ion collisions~\cite{Adamczyk2015,Giacalone2020,Jia2022,Magdy2023}. 
	
	In this work we present the first applications of the SMMC method to actinides. The larger model space and necessity to calculate observables at very large inverse temperatures $\beta = 1/T$ (in order to cool the nucleus to its ground state) present additional technical challenges compared with previous applications of SMMC in lanthanide nuclei. The calculated NLDs are found to be in similarly good agreement with experiment as has been seen in the lighter lanthanides. This work constitutes the first step towards studying statistical properties of actinides that are much beyond the reach of conventional large-scale shell-model calculations. 

\ssec{Shell-model Monte Carlo}  The SMMC method is based on the canonical ensemble describing a nucleus at finite temperature $T$. The Gibbs operator $e^{-\beta \hat{H}}$ (where $\beta=1/T$) is transformed by the Hubbard-Stratonovich (HS) transformation~\cite{Stratonovich1957,Hubbard1959} into a functional integral of one-body propagators $\hat{U}_{\sigma}$ that describe non-interacting nucleons in a set of external, imaginary-time-dependent fields $\sigma(\tau)$ with $0 \le \tau \le \beta$ 
	\begin{equation}
		e^{-\beta \hat{H}} = \int D[\sigma] G_{\sigma} \hat{U}_{\sigma},
	\end{equation} 
where $G_{\sigma}$ is a Gaussian weight and $D[\sigma]$ is the integral measure.

The thermal average of an observable $\hat{O}$ at inverse temperature $\beta$ is then given by
\begin{equation} \label{eq_MC}
	\langle \hat{O} \rangle = \frac{\textrm{Tr}(\hat{O}e^{-\beta \hat{H}})}{\textrm{Tr}(e^{-\beta \hat{H}})} = \frac{\int D[\sigma] G_{\sigma} \langle \hat{O} \rangle_{\sigma} \textrm{Tr} \hat{U}_{\sigma}}{\int D[\sigma] G_{\sigma} \textrm{Tr} \hat{U}_{\sigma}}\;,
\end{equation}
where $\langle \hat{O} \rangle_\sigma = \textrm{Tr} (\hat O \hat U_\sigma) / \textrm{Tr}\, \hat U_\sigma$.
The integration over the auxiliary fields is evaluated using Monte Carlo methods, in which the fields $\sigma(\tau)$ are samples with the positive-definite weight $W_{\sigma} \equiv G_{\sigma} | \textrm{Tr} \hat{U}_{\sigma}|$.  In this work we use good-sign interactions, for which the Monte-Carlo sign $\Phi_{\sigma} \equiv \textrm{Tr} \hat{U}_{\sigma}/| \textrm{Tr} \hat{U}_{\sigma}|$ is always positive in the framework of the grand-canonical ensemble. 

 Quantities in the integrand of Eq.~(\ref{eq_MC}) can be calculated using the matrix representation ${\bf U}_{\sigma}$ of the propagator $\hat{U}_{\sigma}$ in the single-particle model space, reducing the required matrix algebra to matrices the size of the single-particle space ($\sim 100$) rather than the combinatorially-large spaces required in direct diagonalization methods. For example, the grand-canonical trace of
$\hat{U}_{\sigma}$ is given by
\begin{equation}
	\textrm{Tr} \, \hat{U}_{\sigma} = \det (\boldsymbol{1} + {\bf U}_{\sigma}).
\end{equation}

The imaginary-time interval $0 \le \tau \le \beta$ is discretized into $N_{\tau}$ time slices of width $\Delta \beta = \beta/N_{\tau}$. For any given value of $\beta$, thermal averages of observables are calculated for $\Delta \beta = 1/32$ MeV$^{-1}$ and $\Delta \beta = 1/64$ MeV$^{-1}$. The continuous time limit $\Delta \beta \rightarrow 0$ of the expectation value of an observable is then determined by linear extrapolation of the expectation values computed at these two values of $\Delta \beta$.

We work in the canonical ensemble with fixed proton and neutron numbers so the grand-canonical traces in Eq.~(\ref{eq_MC}) are replaced by canonical traces at fixed particle number $A$, i.e., we substitute $\textrm{Tr} \rightarrow \textrm{Tr}_A$. 
This is accomplished by an exact particle-number projection onto $A$ valence nucleons via a discrete Fourier transform~\cite{Ormand1994,Alhassid1999}
 \begin{equation}
	\textrm{Tr}_{A}\hat{U}_{\sigma} = \frac{e^{-\beta \mu A}}{N_s} \sum_{m=1}^{N_s} e^{-i \varphi_m A} \det (\boldsymbol{1} +e^{i \varphi_m + \beta \mu} \bf{U_{\sigma}}) \;,
\end{equation}
where $N_s$ is the number of single-particle states, $\varphi_m=2\pi m/N_s$ are quadrature points and $\mu$ is a real chemical potential introduced to stabilize the Fourier sum.
In practice, we use two particle-number projections onto fixed proton and neutron numbers.

For good-sign interactions (in the grand-canonical ensemble), the projection onto an even number of particles keeps the good sign, but the projection on an odd number of particles leads to a sign problem at low temperatures. This makes direct calculations of the ground-state energy of odd-mass nuclei impossible. Here we use the partition function extrapolation method introduced in Ref.~\cite{Alhassid2024} to calculate the ground-state energy of the odd-mass actinides (see the Supplemental Material~\cite{suppmat}).
	
\ssec{Model space and interaction for actinides}  A CI shell-model Hamiltonian has been developed to describe statistical properties of actinides. The model space consists of spherical orbitals and single-particle energies computed from a central Woods-Saxon potential with spin-orbit coupling. The doubly-magic $^{208}$Pb is used as the inert core. For protons, the valence space consists of the 82-126 shell plus the $1g_{9/2}$ orbital, while the neutron valence space consists of the 126-184 shell plus the $1h_{11/2}$ orbital~\cite{Nakada}. For the heaviest nuclei studied in this work, the total m-scheme model space dimension is greater than $10^{32}$.
	
	The interaction consists of a pairing-plus-multipole interaction which has a good Monte-Carlo sign and includes the dominant collective components of realistic nuclear interactions~\cite{Dufour1996}. It is given by
	\begin{equation}
		 -\sum_{\nu} g_{\nu} \hat{P}_{\nu}^{\dagger} \hat{P}_{\nu} - \sum_{\lambda} \chi_{\lambda} :(\hat{\mathcal{O}}_{\lambda p}+\hat{\mathcal{O}}_{\lambda n}) \cdot (\hat{\mathcal{O}}_{\lambda p}+\hat{\mathcal{O}}_{\lambda n}):,
	\end{equation}
where $\nu = p,n$ is the nucleon species, $\lambda=2,3,4$ corresponds to the $2^{\lambda}$-pole interactions, $P_{\nu}^{\dagger}$ is the monopole pair creation operator, $\hat{\mathcal{O}}_{\lambda \nu}$ are the surface-peaked multipole operators, and $::$ denotes normal ordering. The multipole strengths are defined as $\chi_{\lambda} = \chi k_{\lambda}$ where $\chi$ is determined self-consistently \cite{Alhassid1996} and $k_{\lambda}$ are renormalization factors that take into account the effects of the inert core. The interaction coefficients are parameterized by the following $N$-dependent functional forms:
\begin{equation}
	\begin{gathered}
		g_p = 0.111 + 0.00038(N-148)\\
		g_n = 0.0681 + 0.00047(N-148)^{5/4}\\
		k_2 = 2.299 - \frac{2.39}{(N-148)^2+21.206}-0.0149(N-148)\\
		k_3 = k_4 = 1.0 \\
	\end{gathered}
\end{equation}
These coefficients were determined such that they reproduce known odd-even mass staggerings and models of nuclear state densities fitted to experimental data. See the Supplemental Material~\cite{suppmat} and references therein~\cite{Gilbert1965,Krieger1990,Cwiok1996,Donati2005,Goriely2008,Bonett-Matiz2013} for more details.

\ssec{State and level densities of actinides}  The nuclear state density (NSD) at energy $E$ is the inverse Laplace transform of the nuclear partition function $Z(\beta)$. This transform is numerically ill-behaved, and instead we compute it in the saddle-point approximation where it provides the average NSD
\begin{equation}
	\rho(E) = \frac{1}{2\pi i} \int_{-i\infty}^{+i\infty} d\beta e^{\beta E} Z(\beta) \approx \sqrt{\frac{\beta^2}{2\pi C}} e^{S(E)} \;.
	\label{eq_rho}
\end{equation}
	In Eq.~(\ref{eq_rho}) $S$ and $C$ are, respectively, the canonical entropy and heat capacity.  The logarithm of the partition function $\ln Z(\beta)$ is computed by integrating the thermodynamic relation $ -\partial \ln Z/ \partial \beta=E(\beta)\equiv \langle H\rangle$ and $\beta$ is determined as a function of $E$ using the saddle-point condition $E(\beta)=E$.  The canonical entropy is then given by $S(E) = \ln Z + \beta E$, and the canonical heat capacity $C$ is computed with the method of Ref.~\cite{Liu2001} using correlated errors to reduce statistical errors. Thermal energies $E(\beta)$ are calculated up to $\beta = 40$ MeV$^{-1}$ for even-even nuclei and $\beta = 8$ MeV$^{-1}$ for odd-mass nuclei. The corresponding temperature for the even-even nuclei is sufficiently low so that the ground-state energy $E_0$ can be determined by fitting the thermal energy to a form based on a ground-state rotational band. Thermal energies $E(\beta)$ are calculated up to $\beta = 40$ MeV$^{-1}$ for even-even nuclei and $\beta = 8$ MeV$^{-1}$ for odd-mass nuclei. The corresponding temperature for the odd-mass nuclei is not sufficiently low to directly extract $E_0$ and we use the recently developed partition function extrapolation method~\cite{Alhassid2024}.  The excitation energy is then calculated using $E_x=E-E_0$. 
	 
\begin{figure*}
	\includegraphics[width=\linewidth]{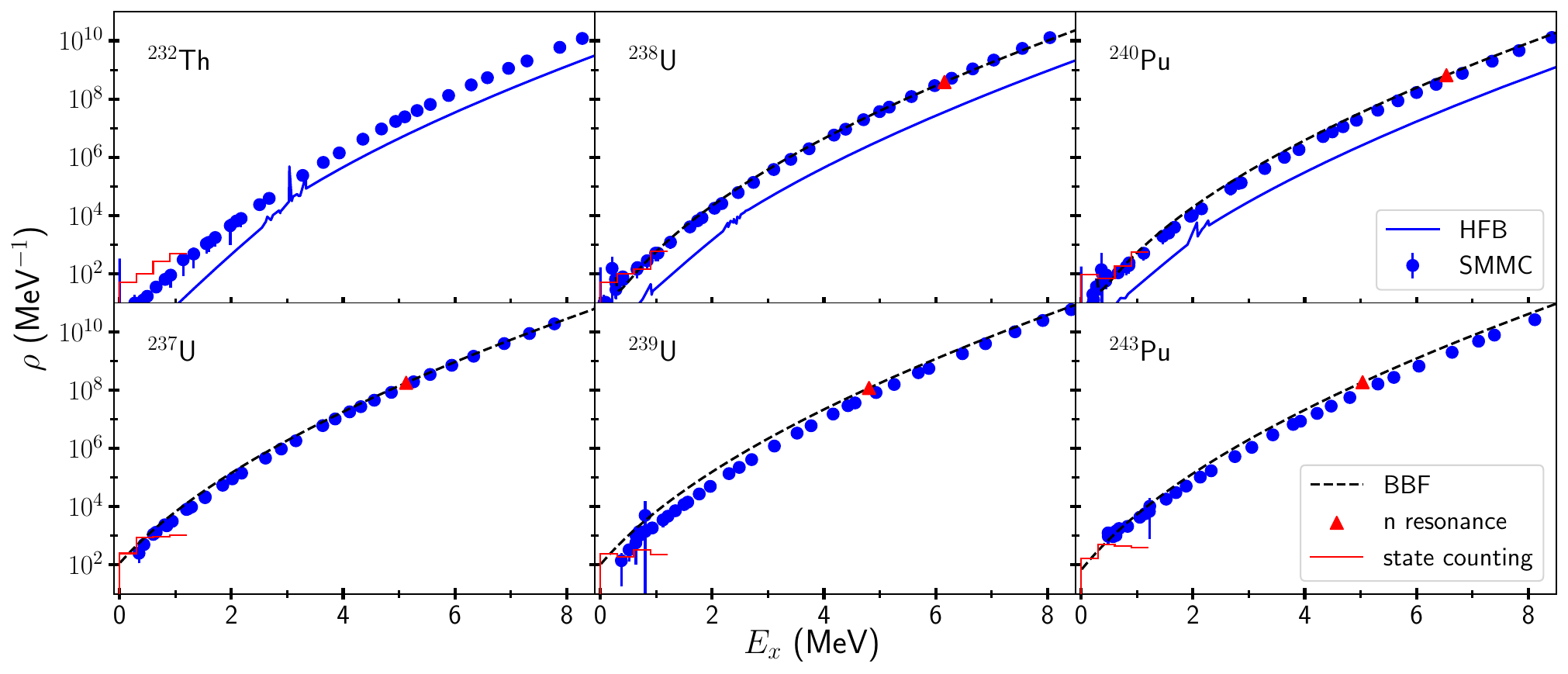}
	\caption{Comparison of the NSDs for actinides calculated in the SMMC (solid circles), HFB (solid lines) and the experimentally determined~\cite{Ozen2025}  BBF (\ref{eq_bbf}) (dashed lines). The red triangles are the NSDs at the neutron separation energies $S_n$ and the histograms describe experimentally known states with assigned spins up to 1.2 MeV \cite{Capote2009}.}
	\label{fig_NSD}
\end{figure*}

	In comparing the SMMC NSDs with experimental data, we used the back-shifted Bethe formula (BBF)~\cite{Dilg1973}
\begin{equation}
	\rho_{\textrm{BBF}}(E_x) = \frac{\sqrt{\pi}e^{2\sqrt{a(E_x-\Delta)}}}{12a^{1/4}(E_x-\Delta)^{5/4}},
	\label{eq_bbf} 
\end{equation}
	where $a$ and $\Delta$ are parameters determined by a fit to low-lying state counting and neutron resonance average spacing data~\cite{Ozen2025} taken from RIPL-3~\cite{Capote2009}. Figure \ref{fig_NSD} compares the NSDs for the actinides calculated with the SMMC (solid circles) and in the finite-temperature Hartree-Fock-Boguliubov (HFB) approximation~\cite{Goodman1981} (solid lines) using the code HF-SHELL\footnote{HF-SHELL only supports calculations in even-even nuclei.}~\cite{Ryssens2021}. While the SMMC shows good agreement with the experimental results, the HFB results are lower by about an order of magnitude. The enhancement of the SMMC NSD relative to the HFB density is attributed to rotational states which are missed by the HFB as these nuclei are strongly deformed ($0.24 < \beta_2 < 0.3$). Previous results have shown that the SMMC and HFB show better agreement with each other as the nuclear deformation decreases~\cite{Guttormsen2021}. 
	
	\begin{figure}
		\includegraphics[width=\linewidth]{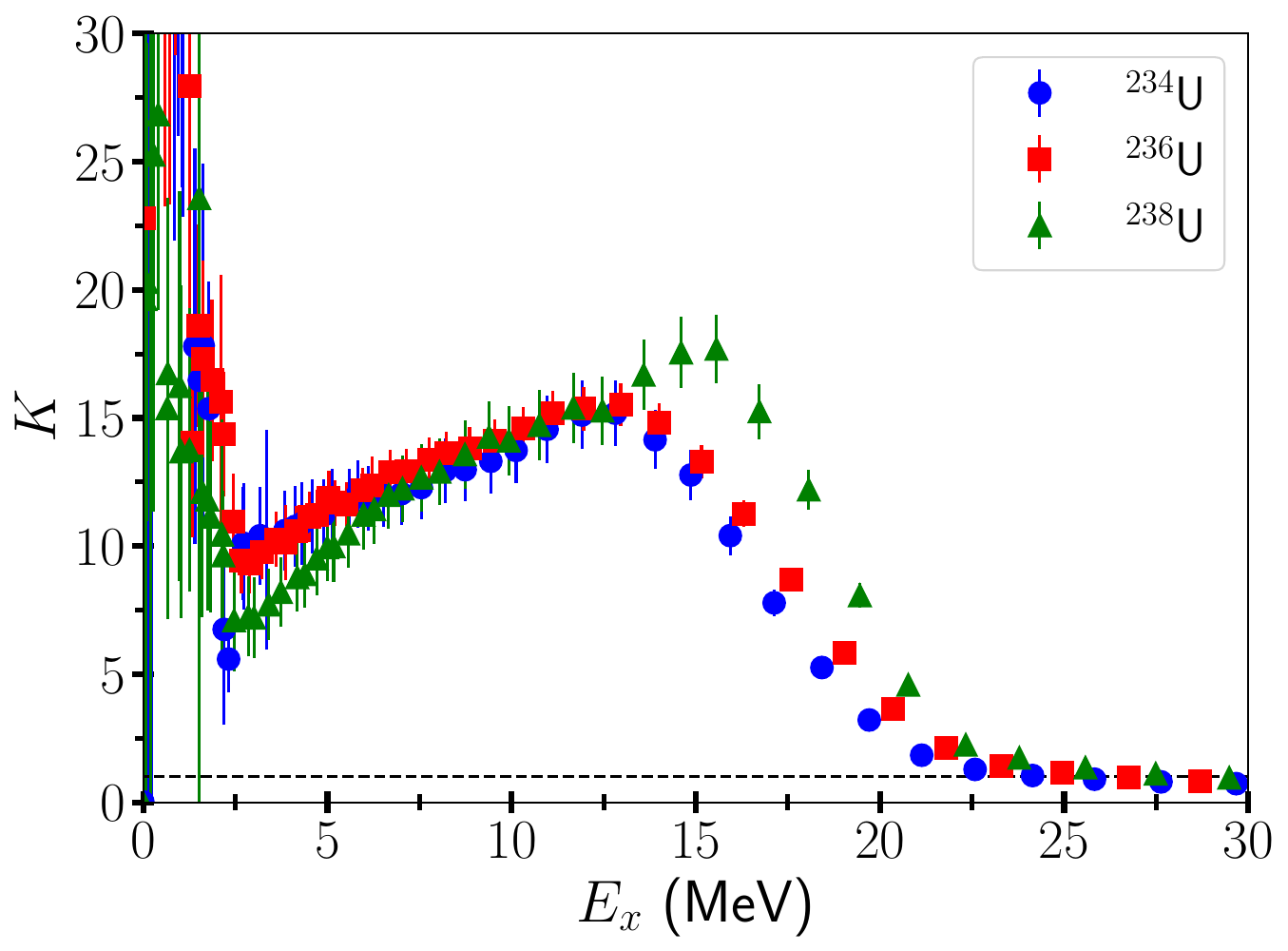}
		\caption{The enhancement factors $K = \rho_{\textrm{SMMC}}/\rho_{\textrm{HFB}}$ of $^{234}$U, $^{236}$U, and $^{238}$U versus excitation energy $E_x$. The dashed line corresponds to $K=1$.}
		\label{fig_K}
	\end{figure}
	
	The contribution of these rotational bands is quantified by the enhancement factor $K = \rho_{\textrm{SMMC}}/\rho_{\textrm{HFB}}$. The energy dependence of $K$, shown in Fig. \ref{fig_K} for the even-even uranium isotopes, displays some important features. For the excitation energies of interest in this work, the enhancement factors are between 10 and 25, indicating a large number of rotational bands that are missed in the HFB (the latter takes into account only the intrinsic band head states). The rapid decrease in $K$ starting around $E_x \sim 15$ MeV is due to the transition from deformed to spherical shapes of the nuclei. At the highest excitation energies shown in Fig.~\ref{fig_K}, when the nuclei are essentially spherical, $K \approx 1$ indicating agreement between the HFB and the SMMC as the rotational bands have disappeared. The large enhancement at very low excitation energies is an unphysical artifact of the non-conservation of particle number in the HFB. 
	
	\begin{figure*}[bth]
		\includegraphics[width=\linewidth]{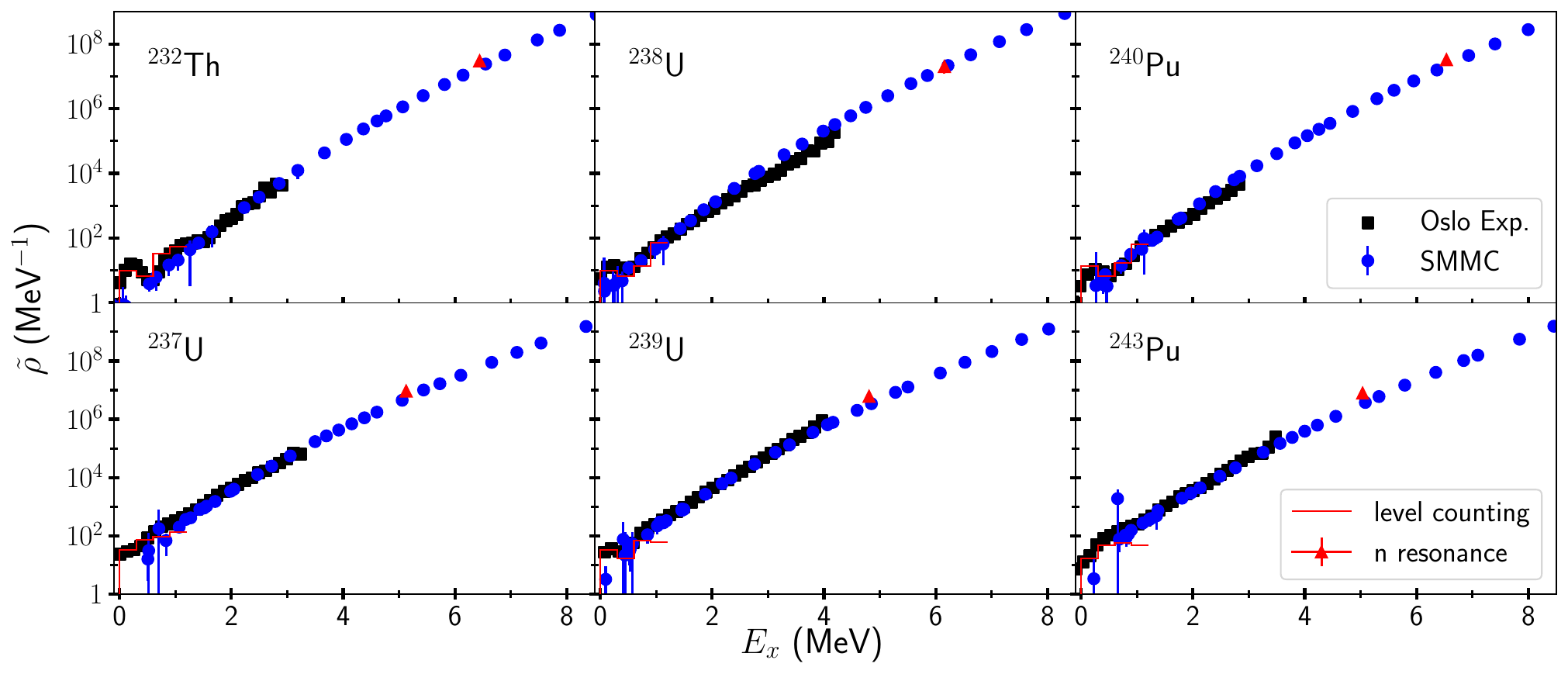}
		\caption{Comparison of the SMMC NLDs (solid circles)  with the Oslo method experiments~\cite{Guttormsen2013,Laplace2016,Zeiser2019} (solid squares) for the same actinides as in Fig.~\ref{fig_NSD}. The red triangles are the NLDs at the neutron separation energies $S_n$ and the histograms are experimentally known levels up to 1.2 MeV~\cite{Capote2009}.}
		\label{fig_NLDs}
	\end{figure*} 

\begin{table*}[bth]
	\begin{tabular}{c|c|c|c|c|c|c|c|c}
		\hline
		\hline
		Nucleus & $S_n$ (MeV) & $J_t$ & \multicolumn{2}{|c|}{$\tilde{\rho}(S_n)$ ($10^6$ MeV$^{-1}$)} & \multicolumn{2}{c}{$\sigma(S_n)$} & \multicolumn{2}{|c}{$D_0$ (eV)} \\
		& & & \multicolumn{2}{c|}{\hspace{0.2cm} Exp. \hspace{0.8cm} SMMC} & \multicolumn{2}{c|}{rigid body \hspace{0.4cm} SMMC} & \multicolumn{2}{c}{\hspace{0.2cm} Exp. \hspace{0.8cm} SMMC} \\
		\hline
		$^{232}$Th & 6.438  & 5/2 & 30. $\pm$ 8.\footnote{The experimental level density for $^{232}$Th was estimated from systematics in Ref. \cite{Guttormsen2013}.} & 20.5 $\pm$ 3.2 & \phantom{rigids}--\phantom{rigids} &8.32 $\pm$ 0.19 & -- & 1.21 $\pm$ 0.20 \\
		$^{237}$U & 5.126 & 0 & 9.3 $\pm$ 1.9 & 5.53 $\pm$ 2.32 &7.95 & 8.05 $\pm$ 0.21 & 14.0 $\pm$ 1.0 & 23.6 $\pm$ 10.0\\
		$^{238}$U & 6.154 & 1/2 & 20. $\pm$ 6. & 19.7 $\pm$ 2.8 & 8.11 & 8.39 $\pm$ 0.17 & 3.5 $\pm$ 0.8  & 3.62 $\pm$ 0.53\\
		$^{239}$U & 4.806 & 0 & 6.1 $\pm$ 1.2 & 3.19 $\pm$ 0.69 & 7.83 & 7.93 $\pm$ 0.22 & 20.3 $\pm$ 0.6 & 39.7 $\pm$ 8.9\\
		$^{240}$Pu & 6.534  & 1/2 & 32.7 $\pm$ 6.6 & 24.7 $\pm$ 3.5 & 8.32 & 8.58 $\pm$ 0.17  & 2.20 $\pm$ 0.09 & 3.01 $\pm$ 0.44\\	
		$^{243}$Pu & 5.034 & 0 & 7.87 $\pm$ 1.93 & 3.54 $\pm$ 2.11 & 7.96 & 8.07 $\pm$ 0.22 & 13.5 $\pm$ 1.5 & 37.1 $\pm$ 22.2\\
		\hline
		\hline
	\end{tabular}
\caption{The SMMC NLDs $\bar \rho(S_n)$ at the neutron separation energies $S_n$ are compared  with the respective NLDs determined from neutron resonance spacing data~\cite{Guttormsen2013,Laplace2016,Zeiser2019}. We also compare the average s-wave neutron resonance spacing $D_0$ calculated with the SMMC (\ref{eq_d0}) with the experimentally measured values~\cite{Capote2009}.}
\label{table_sn}
\end{table*}	

	The NSD counts the total number of states including the $2J+1$ magnetic degeneracy for each level of spin $J$. However, the NLD, measured in experiments such as the Oslo method, counts each level with spin $J$ only once. The NLD can be computed directly through the use of spin projection~\cite{Alhassid2007}. For each level of spin $J$  in an even-mass (odd-mass) nucleus there exists exactly one state with $M=0$ ($M=1/2$). Thus by projecting thermal observables onto $M=0$ ($M=1/2$) for even-mass (odd-mass) nuclei the contribution of only one state from each level is counted~\cite{Alhassid2015}. 
	
        The trace over an observable $\hat{O}$ at fixed spin component $M$ is calculated using a discrete Fourier transform
\begin{equation}
	\textrm{Tr}_M \hat{O} = \frac{1}{2J_s + 1} \sum_{k=-J_s}^{J_s} e^{i \varphi_k M} \textrm{Tr}(e^{i \varphi_k \hat{J}_z}\hat{O}),
	\label{eq_proj}
\end{equation}
	where $\varphi_k =  \frac{k\pi}{J_s + 1/2}$ are quadrature points with $k = -J_s, ... , J_s$ and $J_s$ being the maximum possible spin in the model space. The projected densities $\rho_M$ can be calculated using an equation similar to Eq.~(\ref{eq_rho}) but with the projected thermal energies $E_M$ replacing the thermal energy $E$. The NLDs $\tilde\rho$ are then given by $\tilde\rho =\rho_{M=0}$ ($\tilde\rho =\rho_{M=1/2}$) for even-mass (odd-mass) nuclei~\cite{Alhassid2015}.
The Fourier sum in Eq. (\ref{eq_proj}) can lead to non-positive $M$-projected traces for non-zero $M$, resulting in a sign problem for $M\neq 0$ at very low temperatures.
	Figure \ref{fig_NLDs} shows the NLDs calculated with the SMMC. The SMMC results show good agreement with Oslo method data (solid squares), counting of low-lying levels (red histograms), and neutron resonance data (red triangle) at the neutron separation energy $S_n$. Table \ref{table_sn} provides the calculated and measured values of the NLDs at the neutron separation energies. The NLDs at these energies are needed explicitly by the Oslo method, and are usually estimated via neutron resonance spacing data and the spin-cutoff model.

	The spin (as well as parity) distribution of level densities is another important input in the calculation of nuclear reaction rates~\cite{Rauscher1997}. This spin distribution is also necessary for calculating total level densities from the measured average neutron resonance spacings~\cite{Schiller2000}. Typically, it is assumed that at the neutron resonance energy the spin distribution follows the spin-cutoff model~\cite{Ericson1960}
	\begin{equation}
	\frac{\rho_J(E_x)}{\rho(E_x)} = \frac{2J+1}{2 \sqrt{2 \pi} \sigma^3} e^{-J(J+1)/2\sigma^2},
	\label{eq_sc}
\end{equation}
where $\sigma$ is the spin-cutoff parameter, which depends on $E_x$ and is related to a thermal moment of inertia $I$ by $\sigma^2=IT/\hbar^2$. 

Assuming that both parities have equal densities at $S_n$, the average s-wave neutron resonance spacing $D_0$ of the target nucleus in the neutron resonance experiment (i.e., the nucleus with one neutron fewer than the nucleus whose NLD is being considered) can be related to the spin-dependent level densities by 
\begin{equation}
	\frac{2}{D_0} = \begin{cases}
		{\rho}_{1/2}(S_n) & \textrm{if } J_t = 0 \\
		({\rho}_{J_t-1/2}(S_n) + {\rho}_{J_t+1/2}(S_n)) & \textrm{if } J_t \ne 0,
	\end{cases}
\end{equation}
	where $J_t$ is the ground-state spin of the target nucleus. Further assuming that the spin distribution is well described by the spin-cutoff model (\ref{eq_sc}), we find
\begin{equation}
	\tilde{\rho}(S_n) = \frac{2 \sigma^2}{D_0((J_t+1)e^{-(J_t+1)^2/2\sigma^2} + J_t e^{-J_t^2/2\sigma^2} )}. 
	\label{eq_d0}
\end{equation}

In the SMMC, the spin-projected level densities are calculated from the $M$-projected level densities using $\rho_J= \rho_{M=J} - \rho_{M=J+1}$. 
By computing the spin-projected NLDs, we can determine the spin-distribution $\rho_J/\rho$ at fixed excitation energy $E_x$.  We find that at excitation energies near $S_n$, the spin distribution is very well described by the spin cutoff model. A fit is then performed to determine the spin-cutoff parameter $\sigma(S_n)$ with high accuracy and used to calculate the average neutron resonance $s$-wave spacing $D_0$.

The fitted values of $\sigma(S_n)$ shown in Table \ref{table_sn} are found to be within a few percent of their rigid-body values, as well as from the empirical model of Refs.~\cite{Zhongfu1991,vonEgidy2005} that was used in the Oslo method analyses.  For the even-even nuclei, we find the SMMC results for $\tilde{\rho}(S_n)$ and $D_0$ to agree with their experimental values.  For the odd-mass nuclei, the SMMC values of $\tilde{\rho}(S_n)$ and $D_0$ are, respectively, lower and higher than their corresponding experimental values. We note however that the uncertainties in these quantities for the odd-mass nuclei are much larger due to the larger uncertainties in the  ground-state energies.

\ssec{Conclusion} In conclusion, we have carried out the first microscopic calculations of NSDs and NLDs in the actinide region using the SMMC. The SMMC is the only shell-model method currently able to handle the large model spaces required for this mass region. Our results are in agreement with experimental level counting data at low excitations, neutron resonance data, and recent Oslo method results. 
The SMMC spin projection method enables predictions of the spin cutoff parameter and neutron resonance average spacing in nuclei that have not yet been studied experimentally.
	
\ssec{Acknowledgments}  This work was supported in part by the U.S. DOE grant No. DE-SC0019521. Computational resources used for this work were provided by the National Energy Research Scientific Computing Center (NERSC), a U.S. DOE Office of Science User Facility operated under Contract No. DE-AC02-05CH11231.

The data that support the findings of this article are openly available~\cite{data}.
	
\bibliography{actinideNLDs}

\begin{thebibliography}{54}%
\makeatletter
\providecommand \@ifxundefined [1]{%
 \@ifx{#1\undefined}
}%
\providecommand \@ifnum [1]{%
 \ifnum #1\expandafter \@firstoftwo
 \else \expandafter \@secondoftwo
 \fi
}%
\providecommand \@ifx [1]{%
 \ifx #1\expandafter \@firstoftwo
 \else \expandafter \@secondoftwo
 \fi
}%
\providecommand \natexlab [1]{#1}%
\providecommand \enquote  [1]{``#1''}%
\providecommand \bibnamefont  [1]{#1}%
\providecommand \bibfnamefont [1]{#1}%
\providecommand \citenamefont [1]{#1}%
\providecommand \href@noop [0]{\@secondoftwo}%
\providecommand \href [0]{\begingroup \@sanitize@url \@href}%
\providecommand \@href[1]{\@@startlink{#1}\@@href}%
\providecommand \@@href[1]{\endgroup#1\@@endlink}%
\providecommand \@sanitize@url [0]{\catcode `\\12\catcode `\$12\catcode
  `\&12\catcode `\#12\catcode `\^12\catcode `\_12\catcode `\%12\relax}%
\providecommand \@@startlink[1]{}%
\providecommand \@@endlink[0]{}%
\providecommand \url  [0]{\begingroup\@sanitize@url \@url }%
\providecommand \@url [1]{\endgroup\@href {#1}{\urlprefix }}%
\providecommand \urlprefix  [0]{URL }%
\providecommand \Eprint [0]{\href }%
\providecommand \doibase [0]{https://doi.org/}%
\providecommand \selectlanguage [0]{\@gobble}%
\providecommand \bibinfo  [0]{\@secondoftwo}%
\providecommand \bibfield  [0]{\@secondoftwo}%
\providecommand \translation [1]{[#1]}%
\providecommand \BibitemOpen [0]{}%
\providecommand \bibitemStop [0]{}%
\providecommand \bibitemNoStop [0]{.\EOS\space}%
\providecommand \EOS [0]{\spacefactor3000\relax}%
\providecommand \BibitemShut  [1]{\csname bibitem#1\endcsname}%
\let\auto@bib@innerbib\@empty
\bibitem [{\citenamefont {Hilaire}\ and\ \citenamefont
  {Goriely}(2021)}]{Hilaire2021}%
  \BibitemOpen
  \bibfield  {author} {\bibinfo {author} {\bibfnamefont {S.}~\bibnamefont
  {Hilaire}}\ and\ \bibinfo {author} {\bibfnamefont {S.}~\bibnamefont
  {Goriely}},\ }\bibfield  {title} {\bibinfo {title} {{Towards More Predictive
  Nuclear Reaction Modelling}},\ }\href
  {https://doi.org/10.1007/978-3-030-58082-7_1} {\bibfield  {journal} {\bibinfo
   {journal} {Springer Proc. Phys.}\ }\textbf {\bibinfo {volume} {254}},\
  \bibinfo {pages} {3} (\bibinfo {year} {2021})}\BibitemShut {NoStop}%
\bibitem [{\citenamefont {Arnould}\ \emph {et~al.}(2007)\citenamefont
  {Arnould}, \citenamefont {Goriely},\ and\ \citenamefont
  {Takahashi}}]{Arnould2007}%
  \BibitemOpen
  \bibfield  {author} {\bibinfo {author} {\bibfnamefont {M.}~\bibnamefont
  {Arnould}}, \bibinfo {author} {\bibfnamefont {S.}~\bibnamefont {Goriely}},\
  and\ \bibinfo {author} {\bibfnamefont {K.}~\bibnamefont {Takahashi}},\
  }\bibfield  {title} {\bibinfo {title} {{The r-process of stellar
  nucleosynthesis: Astrophysics and nuclear physics achievements and
  mysteries}},\ }\href {https://doi.org/10.1016/j.physrep.2007.06.002}
  {\bibfield  {journal} {\bibinfo  {journal} {Phys. Rept.}\ }\textbf {\bibinfo
  {volume} {450}},\ \bibinfo {pages} {97} (\bibinfo {year} {2007})}\BibitemShut
  {NoStop}%
\bibitem [{\citenamefont {Drut}\ \emph {et~al.}(2010)\citenamefont {Drut},
  \citenamefont {Furnstahl},\ and\ \citenamefont {Platter}}]{Drut2010}%
  \BibitemOpen
  \bibfield  {author} {\bibinfo {author} {\bibfnamefont {J.~E.}\ \bibnamefont
  {Drut}}, \bibinfo {author} {\bibfnamefont {R.~J.}\ \bibnamefont
  {Furnstahl}},\ and\ \bibinfo {author} {\bibfnamefont {L.}~\bibnamefont
  {Platter}},\ }\bibfield  {title} {\bibinfo {title} {{Toward ab initio density
  functional theory for nuclei}},\ }\href
  {https://doi.org/10.1016/j.ppnp.2009.09.001} {\bibfield  {journal} {\bibinfo
  {journal} {Prog. Part. Nucl. Phys.}\ }\textbf {\bibinfo {volume} {64}},\
  \bibinfo {pages} {120} (\bibinfo {year} {2010})}\BibitemShut {NoStop}%
\bibitem [{\citenamefont {Tsunoda}\ \emph {et~al.}(2017)\citenamefont
  {Tsunoda}, \citenamefont {Otsuka}, \citenamefont {Shimizu}, \citenamefont
  {Hjorth-Jensen}, \citenamefont {Takayanagi},\ and\ \citenamefont
  {Suzuki}}]{Tsunoda2017}%
  \BibitemOpen
  \bibfield  {author} {\bibinfo {author} {\bibfnamefont {N.}~\bibnamefont
  {Tsunoda}}, \bibinfo {author} {\bibfnamefont {T.}~\bibnamefont {Otsuka}},
  \bibinfo {author} {\bibfnamefont {N.}~\bibnamefont {Shimizu}}, \bibinfo
  {author} {\bibfnamefont {M.}~\bibnamefont {Hjorth-Jensen}}, \bibinfo {author}
  {\bibfnamefont {K.}~\bibnamefont {Takayanagi}},\ and\ \bibinfo {author}
  {\bibfnamefont {T.}~\bibnamefont {Suzuki}},\ }\bibfield  {title} {\bibinfo
  {title} {Exotic neutron-rich medium-mass nuclei with realistic nuclear
  forces},\ }\href {https://doi.org/10.1103/PhysRevC.95.021304} {\bibfield
  {journal} {\bibinfo  {journal} {Phys. Rev. C}\ }\textbf {\bibinfo {volume}
  {95}},\ \bibinfo {pages} {021304} (\bibinfo {year} {2017})}\BibitemShut
  {NoStop}%
\bibitem [{\citenamefont {Shimizu}\ \emph {et~al.}(2019)\citenamefont
  {Shimizu}, \citenamefont {Mizusaki}, \citenamefont {Utsuno},\ and\
  \citenamefont {Tsunoda}}]{Shimizu2019}%
  \BibitemOpen
  \bibfield  {author} {\bibinfo {author} {\bibfnamefont {N.}~\bibnamefont
  {Shimizu}}, \bibinfo {author} {\bibfnamefont {T.}~\bibnamefont {Mizusaki}},
  \bibinfo {author} {\bibfnamefont {Y.}~\bibnamefont {Utsuno}},\ and\ \bibinfo
  {author} {\bibfnamefont {Y.}~\bibnamefont {Tsunoda}},\ }\bibfield  {title}
  {\bibinfo {title} {{Thick-Restart Block Lanczos Method for Large-Scale
  Shell-Model Calculations}},\ }\href
  {https://doi.org/10.1016/j.cpc.2019.06.011} {\bibfield  {journal} {\bibinfo
  {journal} {Comput. Phys. Commun.}\ }\textbf {\bibinfo {volume} {244}},\
  \bibinfo {pages} {372} (\bibinfo {year} {2019})}\BibitemShut {NoStop}%
\bibitem [{\citenamefont {Alhassid}(2017)}]{Alhassid2017}%
  \BibitemOpen
  \bibfield  {author} {\bibinfo {author} {\bibfnamefont {Y.}~\bibnamefont
  {Alhassid}},\ }\bibfield  {title} {\bibinfo {title} {Auxiliary-field quantum
  monte carlo methods in nuclei},\ }in\ \href@noop {} {\emph {\bibinfo
  {booktitle} {Emergent Phenomena in Atomic Nuclei from Large-Scale Modeling: a
  Symmetry-Guided Perspective}}},\ \bibinfo {editor} {edited by\ \bibinfo
  {editor} {\bibfnamefont {K.~D.}\ \bibnamefont {Launey}}}\ (\bibinfo
  {publisher} {World Scientific},\ \bibinfo {address} {Singapore},\ \bibinfo
  {year} {2017})\ pp.\ \bibinfo {pages} {267--298}\BibitemShut {NoStop}%
\bibitem [{\citenamefont {Alhassid}\ \emph {et~al.}(2008)\citenamefont
  {Alhassid}, \citenamefont {Fang},\ and\ \citenamefont
  {Nakada}}]{Alhassid2008}%
  \BibitemOpen
  \bibfield  {author} {\bibinfo {author} {\bibfnamefont {Y.}~\bibnamefont
  {Alhassid}}, \bibinfo {author} {\bibfnamefont {L.}~\bibnamefont {Fang}},\
  and\ \bibinfo {author} {\bibfnamefont {H.}~\bibnamefont {Nakada}},\
  }\bibfield  {title} {\bibinfo {title} {{Heavy deformed nuclei in the shell
  model Monte Carlo method}},\ }\href
  {https://doi.org/10.1103/PhysRevLett.101.082501} {\bibfield  {journal}
  {\bibinfo  {journal} {Phys. Rev. Lett.}\ }\textbf {\bibinfo {volume} {101}},\
  \bibinfo {pages} {082501} (\bibinfo {year} {2008})}\BibitemShut {NoStop}%
\bibitem [{\citenamefont {\"Ozen}\ \emph {et~al.}(2013)\citenamefont {\"Ozen},
  \citenamefont {Alhassid},\ and\ \citenamefont {Nakada}}]{Ozen2013}%
  \BibitemOpen
  \bibfield  {author} {\bibinfo {author} {\bibfnamefont {C.}~\bibnamefont
  {\"Ozen}}, \bibinfo {author} {\bibfnamefont {Y.}~\bibnamefont {Alhassid}},\
  and\ \bibinfo {author} {\bibfnamefont {H.}~\bibnamefont {Nakada}},\
  }\bibfield  {title} {\bibinfo {title} {{Crossover from Vibrational to
  Rotational Collectivity in Heavy Nuclei in the Shell-Model Monte Carlo
  Approach}},\ }\href {https://doi.org/10.1103/PhysRevLett.110.042502}
  {\bibfield  {journal} {\bibinfo  {journal} {Phys. Rev. Lett.}\ }\textbf
  {\bibinfo {volume} {110}},\ \bibinfo {pages} {042502} (\bibinfo {year}
  {2013})}\BibitemShut {NoStop}%
\bibitem [{\citenamefont {\"Ozen}\ \emph {et~al.}(2015)\citenamefont {\"Ozen},
  \citenamefont {Alhassid},\ and\ \citenamefont {Nakada}}]{Ozen2015}%
  \BibitemOpen
  \bibfield  {author} {\bibinfo {author} {\bibfnamefont {C.}~\bibnamefont
  {\"Ozen}}, \bibinfo {author} {\bibfnamefont {Y.}~\bibnamefont {Alhassid}},\
  and\ \bibinfo {author} {\bibfnamefont {H.}~\bibnamefont {Nakada}},\
  }\bibfield  {title} {\bibinfo {title} {{Nuclear state densities of odd-mass
  heavy nuclei in the shell model Monte Carlo approach}},\ }\href
  {https://doi.org/10.1103/PhysRevC.91.034329} {\bibfield  {journal} {\bibinfo
  {journal} {Phys. Rev. C}\ }\textbf {\bibinfo {volume} {91}},\ \bibinfo
  {pages} {034329} (\bibinfo {year} {2015})}\BibitemShut {NoStop}%
\bibitem [{\citenamefont {Alhassid}\ \emph {et~al.}(2014)\citenamefont
  {Alhassid}, \citenamefont {Gilbreth},\ and\ \citenamefont
  {Bertsch}}]{Alhassid2014}%
  \BibitemOpen
  \bibfield  {author} {\bibinfo {author} {\bibfnamefont {Y.}~\bibnamefont
  {Alhassid}}, \bibinfo {author} {\bibfnamefont {C.~N.}\ \bibnamefont
  {Gilbreth}},\ and\ \bibinfo {author} {\bibfnamefont {G.~F.}\ \bibnamefont
  {Bertsch}},\ }\bibfield  {title} {\bibinfo {title} {{Nuclear deformation at
  finite temperature}},\ }\href
  {https://doi.org/10.1103/PhysRevLett.113.262503} {\bibfield  {journal}
  {\bibinfo  {journal} {Phys. Rev. Lett.}\ }\textbf {\bibinfo {volume} {113}},\
  \bibinfo {pages} {262503} (\bibinfo {year} {2014})}\BibitemShut {NoStop}%
\bibitem [{\citenamefont {Guttormsen}\ \emph {et~al.}(2021)\citenamefont
  {Guttormsen}, \citenamefont {Alhassid}, \citenamefont {Ryssens},
  \citenamefont {Ay}, \citenamefont {Ozgur}, \citenamefont {Algin},
  \citenamefont {Larsen}, \citenamefont {Bello~Garrote}, \citenamefont
  {Crespo~Campo}, \citenamefont {Dahl-Jacobsen} \emph
  {et~al.}}]{Guttormsen2021}%
  \BibitemOpen
  \bibfield  {author} {\bibinfo {author} {\bibfnamefont {M.}~\bibnamefont
  {Guttormsen}}, \bibinfo {author} {\bibfnamefont {Y.}~\bibnamefont
  {Alhassid}}, \bibinfo {author} {\bibfnamefont {W.}~\bibnamefont {Ryssens}},
  \bibinfo {author} {\bibfnamefont {K.~O.}\ \bibnamefont {Ay}}, \bibinfo
  {author} {\bibfnamefont {M.}~\bibnamefont {Ozgur}}, \bibinfo {author}
  {\bibfnamefont {E.}~\bibnamefont {Algin}}, \bibinfo {author} {\bibfnamefont
  {A.~C.}\ \bibnamefont {Larsen}}, \bibinfo {author} {\bibfnamefont {F.~L.}\
  \bibnamefont {Bello~Garrote}}, \bibinfo {author} {\bibfnamefont
  {L.}~\bibnamefont {Crespo~Campo}}, \bibinfo {author} {\bibfnamefont
  {T.}~\bibnamefont {Dahl-Jacobsen}}, \emph {et~al.},\ }\bibfield  {title}
  {\bibinfo {title} {{Strong enhancement of level densities in the crossover
  from spherical to deformed neodymium isotopes}},\ }\href
  {https://doi.org/10.1016/j.physletb.2021.136206} {\bibfield  {journal}
  {\bibinfo  {journal} {Phys. Lett. B}\ }\textbf {\bibinfo {volume} {816}},\
  \bibinfo {pages} {136206} (\bibinfo {year} {2021})}\BibitemShut {NoStop}%
\bibitem [{\citenamefont {Mercenne}\ \emph {et~al.}(2024)\citenamefont
  {Mercenne}, \citenamefont {Fanto}, \citenamefont {Ryssens},\ and\
  \citenamefont {Alhassid}}]{Mercenne2024}%
  \BibitemOpen
  \bibfield  {author} {\bibinfo {author} {\bibfnamefont {A.}~\bibnamefont
  {Mercenne}}, \bibinfo {author} {\bibfnamefont {P.}~\bibnamefont {Fanto}},
  \bibinfo {author} {\bibfnamefont {W.}~\bibnamefont {Ryssens}},\ and\ \bibinfo
  {author} {\bibfnamefont {Y.}~\bibnamefont {Alhassid}},\ }\bibfield  {title}
  {\bibinfo {title} {{Magnetic dipole \ensuremath{\gamma}-ray strength
  functions in the crossover from spherical to deformed neodymium isotopes}},\
  }\href {https://doi.org/10.1103/PhysRevC.110.054313} {\bibfield  {journal}
  {\bibinfo  {journal} {Phys. Rev. C}\ }\textbf {\bibinfo {volume} {110}},\
  \bibinfo {pages} {054313} (\bibinfo {year} {2024})}\BibitemShut {NoStop}%
\bibitem [{\citenamefont {DeMartini}\ and\ \citenamefont
  {Alhassid}(2025)}]{DeMartini2025}%
  \BibitemOpen
  \bibfield  {author} {\bibinfo {author} {\bibfnamefont {D.}~\bibnamefont
  {DeMartini}}\ and\ \bibinfo {author} {\bibfnamefont {Y.}~\bibnamefont
  {Alhassid}},\ }\bibfield  {title} {\bibinfo {title} {{Low-energy enhancement
  of the magnetic dipole radiation in odd-mass lanthanides}},\ }\href
  {https://doi.org/10.1103/PhysRevC.111.034315} {\bibfield  {journal} {\bibinfo
   {journal} {Phys. Rev. C}\ }\textbf {\bibinfo {volume} {111}},\ \bibinfo
  {pages} {034315} (\bibinfo {year} {2025})}\BibitemShut {NoStop}%
\bibitem [{\citenamefont {Schiller}\ \emph {et~al.}(2000)\citenamefont
  {Schiller}, \citenamefont {Bergholt}, \citenamefont {Guttormsen},
  \citenamefont {Melby}, \citenamefont {Rekstad},\ and\ \citenamefont
  {Siem}}]{Schiller2000}%
  \BibitemOpen
  \bibfield  {author} {\bibinfo {author} {\bibfnamefont {A.}~\bibnamefont
  {Schiller}}, \bibinfo {author} {\bibfnamefont {L.}~\bibnamefont {Bergholt}},
  \bibinfo {author} {\bibfnamefont {M.}~\bibnamefont {Guttormsen}}, \bibinfo
  {author} {\bibfnamefont {E.}~\bibnamefont {Melby}}, \bibinfo {author}
  {\bibfnamefont {J.}~\bibnamefont {Rekstad}},\ and\ \bibinfo {author}
  {\bibfnamefont {S.}~\bibnamefont {Siem}},\ }\bibfield  {title} {\bibinfo
  {title} {{Extraction of level density and gamma strength function from
  primary gamma spectra}},\ }\href
  {https://doi.org/10.1016/S0168-9002(99)01187-0} {\bibfield  {journal}
  {\bibinfo  {journal} {Nucl. Instrum. Meth. A}\ }\textbf {\bibinfo {volume}
  {447}},\ \bibinfo {pages} {494} (\bibinfo {year} {2000})}\BibitemShut
  {NoStop}%
\bibitem [{\citenamefont {Guttormsen}\ \emph {et~al.}(2005)\citenamefont
  {Guttormsen}, \citenamefont {Chankova}, \citenamefont {Agvaanluvsan},
  \citenamefont {Algin}, \citenamefont {Bernstein}, \citenamefont
  {Ingebretsen}, \citenamefont {Lonnroth}, \citenamefont {Messelt},
  \citenamefont {Mitchell}, \citenamefont {Rekstad} \emph
  {et~al.}}]{Guttormsen2005}%
  \BibitemOpen
  \bibfield  {author} {\bibinfo {author} {\bibfnamefont {M.}~\bibnamefont
  {Guttormsen}}, \bibinfo {author} {\bibfnamefont {R.}~\bibnamefont
  {Chankova}}, \bibinfo {author} {\bibfnamefont {U.}~\bibnamefont
  {Agvaanluvsan}}, \bibinfo {author} {\bibfnamefont {E.}~\bibnamefont {Algin}},
  \bibinfo {author} {\bibfnamefont {L.~A.}\ \bibnamefont {Bernstein}}, \bibinfo
  {author} {\bibfnamefont {F.}~\bibnamefont {Ingebretsen}}, \bibinfo {author}
  {\bibfnamefont {T.}~\bibnamefont {Lonnroth}}, \bibinfo {author}
  {\bibfnamefont {S.}~\bibnamefont {Messelt}}, \bibinfo {author} {\bibfnamefont
  {G.~E.}\ \bibnamefont {Mitchell}}, \bibinfo {author} {\bibfnamefont
  {J.}~\bibnamefont {Rekstad}}, \emph {et~al.},\ }\bibfield  {title} {\bibinfo
  {title} {{Radiative strength functions in Mo-93, Mo-94, Mo-95, Mo-96, Mo-97,
  Mo-98}},\ }\href {https://doi.org/10.1103/PhysRevC.71.044307} {\bibfield
  {journal} {\bibinfo  {journal} {Phys. Rev. C}\ }\textbf {\bibinfo {volume}
  {71}},\ \bibinfo {pages} {044307} (\bibinfo {year} {2005})}\BibitemShut
  {NoStop}%
\bibitem [{\citenamefont {Larsen}\ \emph {et~al.}(2011)\citenamefont {Larsen},
  \citenamefont {Guttormsen}, \citenamefont {Krticka}, \citenamefont {Betak},
  \citenamefont {Burger}, \citenamefont {Gorgen}, \citenamefont {Nyhus},
  \citenamefont {Rekstad}, \citenamefont {Schiller}, \citenamefont {Siem} \emph
  {et~al.}}]{Larsen2011}%
  \BibitemOpen
  \bibfield  {author} {\bibinfo {author} {\bibfnamefont {A.~C.}\ \bibnamefont
  {Larsen}}, \bibinfo {author} {\bibfnamefont {M.}~\bibnamefont {Guttormsen}},
  \bibinfo {author} {\bibfnamefont {M.}~\bibnamefont {Krticka}}, \bibinfo
  {author} {\bibfnamefont {E.}~\bibnamefont {Betak}}, \bibinfo {author}
  {\bibfnamefont {A.}~\bibnamefont {Burger}}, \bibinfo {author} {\bibfnamefont
  {A.}~\bibnamefont {Gorgen}}, \bibinfo {author} {\bibfnamefont {H.~T.}\
  \bibnamefont {Nyhus}}, \bibinfo {author} {\bibfnamefont {J.}~\bibnamefont
  {Rekstad}}, \bibinfo {author} {\bibfnamefont {A.}~\bibnamefont {Schiller}},
  \bibinfo {author} {\bibfnamefont {S.}~\bibnamefont {Siem}}, \emph {et~al.},\
  }\bibfield  {title} {\bibinfo {title} {{Analysis of possible systematic
  errors in the Oslo method}},\ }\href
  {https://doi.org/10.1103/PhysRevC.83.034315} {\bibfield  {journal} {\bibinfo
  {journal} {Phys. Rev. C}\ }\textbf {\bibinfo {volume} {83}},\ \bibinfo
  {pages} {034315} (\bibinfo {year} {2011})},\ \bibinfo {note} {[Erratum:
  Phys.Rev.C 97, 049901 (2018)]}\BibitemShut {NoStop}%
\bibitem [{\citenamefont {Guttormsen}\ \emph {et~al.}(2013)\citenamefont
  {Guttormsen}, \citenamefont {Jurado}, \citenamefont {Wilson}, \citenamefont
  {Aiche}, \citenamefont {Bernstein}, \citenamefont {Ducasse}, \citenamefont
  {Giacoppo}, \citenamefont {Gorgen}, \citenamefont {Gunsing}, \citenamefont
  {Hagen} \emph {et~al.}}]{Guttormsen2013}%
  \BibitemOpen
  \bibfield  {author} {\bibinfo {author} {\bibfnamefont {M.}~\bibnamefont
  {Guttormsen}}, \bibinfo {author} {\bibfnamefont {B.}~\bibnamefont {Jurado}},
  \bibinfo {author} {\bibfnamefont {J.~N.}\ \bibnamefont {Wilson}}, \bibinfo
  {author} {\bibfnamefont {M.}~\bibnamefont {Aiche}}, \bibinfo {author}
  {\bibfnamefont {L.~A.}\ \bibnamefont {Bernstein}}, \bibinfo {author}
  {\bibfnamefont {Q.}~\bibnamefont {Ducasse}}, \bibinfo {author} {\bibfnamefont
  {F.}~\bibnamefont {Giacoppo}}, \bibinfo {author} {\bibfnamefont
  {A.}~\bibnamefont {Gorgen}}, \bibinfo {author} {\bibfnamefont
  {F.}~\bibnamefont {Gunsing}}, \bibinfo {author} {\bibfnamefont {T.~W.}\
  \bibnamefont {Hagen}}, \emph {et~al.},\ }\bibfield  {title} {\bibinfo {title}
  {{Constant-temperature level densities in the quasicontinuum of Th and U
  isotopes}},\ }\href {https://doi.org/10.1103/PhysRevC.88.024307} {\bibfield
  {journal} {\bibinfo  {journal} {Phys. Rev. C}\ }\textbf {\bibinfo {volume}
  {88}},\ \bibinfo {pages} {024307} (\bibinfo {year} {2013})}\BibitemShut
  {NoStop}%
\bibitem [{\citenamefont {Tornyi}\ \emph {et~al.}(2014)\citenamefont {Tornyi},
  \citenamefont {Guttormsen}, \citenamefont {Eriksen}, \citenamefont
  {G\"orgen}, \citenamefont {Giacoppo}, \citenamefont {Hagen}, \citenamefont
  {Krasznahorkay}, \citenamefont {Larsen}, \citenamefont {Renstr\o{}m},
  \citenamefont {Rose} \emph {et~al.}}]{Tornyi2014}%
  \BibitemOpen
  \bibfield  {author} {\bibinfo {author} {\bibfnamefont {T.~G.}\ \bibnamefont
  {Tornyi}}, \bibinfo {author} {\bibfnamefont {M.}~\bibnamefont {Guttormsen}},
  \bibinfo {author} {\bibfnamefont {T.~K.}\ \bibnamefont {Eriksen}}, \bibinfo
  {author} {\bibfnamefont {G.}~\bibnamefont {G\"orgen}}, \bibinfo {author}
  {\bibfnamefont {F.}~\bibnamefont {Giacoppo}}, \bibinfo {author}
  {\bibfnamefont {T.~W.}\ \bibnamefont {Hagen}}, \bibinfo {author}
  {\bibfnamefont {A.}~\bibnamefont {Krasznahorkay}}, \bibinfo {author}
  {\bibfnamefont {A.~C.}\ \bibnamefont {Larsen}}, \bibinfo {author}
  {\bibfnamefont {T.}~\bibnamefont {Renstr\o{}m}}, \bibinfo {author}
  {\bibfnamefont {S.~J.}\ \bibnamefont {Rose}}, \emph {et~al.},\ }\bibfield
  {title} {\bibinfo {title} {{Level density and \ensuremath{\gamma} -ray
  strength function in the odd-odd Np238 nucleus}},\ }\href
  {https://doi.org/10.1103/PhysRevC.89.044323} {\bibfield  {journal} {\bibinfo
  {journal} {Phys. Rev. C}\ }\textbf {\bibinfo {volume} {89}},\ \bibinfo
  {pages} {044323} (\bibinfo {year} {2014})}\BibitemShut {NoStop}%
\bibitem [{\citenamefont {Laplace}\ \emph {et~al.}(2016)\citenamefont
  {Laplace}, \citenamefont {Zeiser}, \citenamefont {Guttormsen}, \citenamefont
  {Larsen}, \citenamefont {Bleuel}, \citenamefont {Bernstein}, \citenamefont
  {Goldblum}, \citenamefont {Siem}, \citenamefont {Bello~Garotte},
  \citenamefont {Brown} \emph {et~al.}}]{Laplace2016}%
  \BibitemOpen
  \bibfield  {author} {\bibinfo {author} {\bibfnamefont {T.~A.}\ \bibnamefont
  {Laplace}}, \bibinfo {author} {\bibfnamefont {F.}~\bibnamefont {Zeiser}},
  \bibinfo {author} {\bibfnamefont {M.}~\bibnamefont {Guttormsen}}, \bibinfo
  {author} {\bibfnamefont {A.~C.}\ \bibnamefont {Larsen}}, \bibinfo {author}
  {\bibfnamefont {D.~L.}\ \bibnamefont {Bleuel}}, \bibinfo {author}
  {\bibfnamefont {L.~A.}\ \bibnamefont {Bernstein}}, \bibinfo {author}
  {\bibfnamefont {B.~L.}\ \bibnamefont {Goldblum}}, \bibinfo {author}
  {\bibfnamefont {S.}~\bibnamefont {Siem}}, \bibinfo {author} {\bibfnamefont
  {F.~L.}\ \bibnamefont {Bello~Garotte}}, \bibinfo {author} {\bibfnamefont
  {J.~A.}\ \bibnamefont {Brown}}, \emph {et~al.},\ }\bibfield  {title}
  {\bibinfo {title} {Statistical properties of $^{243}\text{Pu}$, and
  $^{242}\text{Pu}(n,\ensuremath{\gamma})$ cross section calculation},\ }\href
  {https://doi.org/10.1103/PhysRevC.93.014323} {\bibfield  {journal} {\bibinfo
  {journal} {Phys. Rev. C}\ }\textbf {\bibinfo {volume} {93}},\ \bibinfo
  {pages} {014323} (\bibinfo {year} {2016})}\BibitemShut {NoStop}%
\bibitem [{\citenamefont {Zeiser}\ \emph {et~al.}(2019)\citenamefont {Zeiser},
  \citenamefont {Tveten}, \citenamefont {Potel}, \citenamefont {Larsen},
  \citenamefont {Guttormsen}, \citenamefont {Laplace}, \citenamefont {Siem},
  \citenamefont {Bleuel}, \citenamefont {Goldblum}, \citenamefont {Bernstein}
  \emph {et~al.}}]{Zeiser2019}%
  \BibitemOpen
  \bibfield  {author} {\bibinfo {author} {\bibfnamefont {F.}~\bibnamefont
  {Zeiser}}, \bibinfo {author} {\bibfnamefont {G.~M.}\ \bibnamefont {Tveten}},
  \bibinfo {author} {\bibfnamefont {G.}~\bibnamefont {Potel}}, \bibinfo
  {author} {\bibfnamefont {A.~C.}\ \bibnamefont {Larsen}}, \bibinfo {author}
  {\bibfnamefont {M.}~\bibnamefont {Guttormsen}}, \bibinfo {author}
  {\bibfnamefont {T.~A.}\ \bibnamefont {Laplace}}, \bibinfo {author}
  {\bibfnamefont {S.}~\bibnamefont {Siem}}, \bibinfo {author} {\bibfnamefont
  {D.~L.}\ \bibnamefont {Bleuel}}, \bibinfo {author} {\bibfnamefont {B.~L.}\
  \bibnamefont {Goldblum}}, \bibinfo {author} {\bibfnamefont {L.~A.}\
  \bibnamefont {Bernstein}}, \emph {et~al.},\ }\bibfield  {title} {\bibinfo
  {title} {Restricted spin-range correction in the oslo method: The example of
  nuclear level density and $\ensuremath{\gamma}$-ray strength function from
  $^{239}\mathrm{Pu}(d,p\ensuremath{\gamma})^{240}\mathrm{Pu}$},\ }\href
  {https://doi.org/10.1103/PhysRevC.100.024305} {\bibfield  {journal} {\bibinfo
   {journal} {Phys. Rev. C}\ }\textbf {\bibinfo {volume} {100}},\ \bibinfo
  {pages} {024305} (\bibinfo {year} {2019})}\BibitemShut {NoStop}%
\bibitem [{\citenamefont {Mazurek}\ \emph {et~al.}(2011)\citenamefont
  {Mazurek}, \citenamefont {Schmitt}, \citenamefont {Wieleczko}, \citenamefont
  {Nadtochy},\ and\ \citenamefont {Ademard}}]{Mazurek2011}%
  \BibitemOpen
  \bibfield  {author} {\bibinfo {author} {\bibfnamefont {K.}~\bibnamefont
  {Mazurek}}, \bibinfo {author} {\bibfnamefont {C.}~\bibnamefont {Schmitt}},
  \bibinfo {author} {\bibfnamefont {J.~P.}\ \bibnamefont {Wieleczko}}, \bibinfo
  {author} {\bibfnamefont {P.~N.}\ \bibnamefont {Nadtochy}},\ and\ \bibinfo
  {author} {\bibfnamefont {G.}~\bibnamefont {Ademard}},\ }\bibfield  {title}
  {\bibinfo {title} {{Critical insight into the influence of the potential
  energy surface on fission dynamics}},\ }\href
  {https://doi.org/10.1103/PhysRevC.84.014610} {\bibfield  {journal} {\bibinfo
  {journal} {Phys. Rev. C}\ }\textbf {\bibinfo {volume} {84}},\ \bibinfo
  {pages} {014610} (\bibinfo {year} {2011})}\BibitemShut {NoStop}%
\bibitem [{\citenamefont {Ward}\ \emph {et~al.}(2017)\citenamefont {Ward},
  \citenamefont {Carlsson}, \citenamefont {D\o{}ssing}, \citenamefont
  {M\"oller}, \citenamefont {Randrup},\ and\ \citenamefont
  {\AA{}berg}}]{Ward2017}%
  \BibitemOpen
  \bibfield  {author} {\bibinfo {author} {\bibfnamefont {D.~E.}\ \bibnamefont
  {Ward}}, \bibinfo {author} {\bibfnamefont {B.~G.}\ \bibnamefont {Carlsson}},
  \bibinfo {author} {\bibfnamefont {T.}~\bibnamefont {D\o{}ssing}}, \bibinfo
  {author} {\bibfnamefont {P.}~\bibnamefont {M\"oller}}, \bibinfo {author}
  {\bibfnamefont {J.}~\bibnamefont {Randrup}},\ and\ \bibinfo {author}
  {\bibfnamefont {S.}~\bibnamefont {\AA{}berg}},\ }\bibfield  {title} {\bibinfo
  {title} {Nuclear shape evolution based on microscopic level densities},\
  }\href {https://doi.org/10.1103/PhysRevC.95.024618} {\bibfield  {journal}
  {\bibinfo  {journal} {Phys. Rev. C}\ }\textbf {\bibinfo {volume} {95}},\
  \bibinfo {pages} {024618} (\bibinfo {year} {2017})}\BibitemShut {NoStop}%
\bibitem [{\citenamefont {Adamczyk}\ \emph {et~al.}(2015)\citenamefont
  {Adamczyk} \emph {et~al.}}]{Adamczyk2015}%
  \BibitemOpen
  \bibfield  {author} {\bibinfo {author} {\bibfnamefont {L.}~\bibnamefont
  {Adamczyk}} \emph {et~al.} (\bibinfo {collaboration} {STAR Collaboration}),\
  }\bibfield  {title} {\bibinfo {title} {Azimuthal anisotropy in
  $\mathrm{U}+\mathrm{U}$ and $\mathrm{Au}+\mathrm{Au}$ collisions at rhic},\
  }\href {https://doi.org/10.1103/PhysRevLett.115.222301} {\bibfield  {journal}
  {\bibinfo  {journal} {Phys. Rev. Lett.}\ }\textbf {\bibinfo {volume} {115}},\
  \bibinfo {pages} {222301} (\bibinfo {year} {2015})}\BibitemShut {NoStop}%
\bibitem [{\citenamefont {Giacalone}(2020)}]{Giacalone2020}%
  \BibitemOpen
  \bibfield  {author} {\bibinfo {author} {\bibfnamefont {G.}~\bibnamefont
  {Giacalone}},\ }\bibfield  {title} {\bibinfo {title} {{Constraining the
  quadrupole deformation of atomic nuclei with relativistic nuclear
  collisions}},\ }\href {https://doi.org/10.1103/PhysRevC.102.024901}
  {\bibfield  {journal} {\bibinfo  {journal} {Phys. Rev. C}\ }\textbf {\bibinfo
  {volume} {102}},\ \bibinfo {pages} {024901} (\bibinfo {year}
  {2020})}\BibitemShut {NoStop}%
\bibitem [{\citenamefont {Jia}\ \emph {et~al.}(2022)\citenamefont {Jia},
  \citenamefont {Huang},\ and\ \citenamefont {Zhang}}]{Jia2022}%
  \BibitemOpen
  \bibfield  {author} {\bibinfo {author} {\bibfnamefont {J.}~\bibnamefont
  {Jia}}, \bibinfo {author} {\bibfnamefont {S.}~\bibnamefont {Huang}},\ and\
  \bibinfo {author} {\bibfnamefont {C.}~\bibnamefont {Zhang}},\ }\bibfield
  {title} {\bibinfo {title} {{Probing nuclear quadrupole deformation from
  correlation of elliptic flow and transverse momentum in heavy ion
  collisions}},\ }\href {https://doi.org/10.1103/PhysRevC.105.014906}
  {\bibfield  {journal} {\bibinfo  {journal} {Phys. Rev. C}\ }\textbf {\bibinfo
  {volume} {105}},\ \bibinfo {pages} {014906} (\bibinfo {year}
  {2022})}\BibitemShut {NoStop}%
\bibitem [{\citenamefont {Magdy}(2023)}]{Magdy2023}%
  \BibitemOpen
  \bibfield  {author} {\bibinfo {author} {\bibfnamefont {N.}~\bibnamefont
  {Magdy}},\ }\bibfield  {title} {\bibinfo {title} {{Impact of nuclear
  deformation on collective flow observables in relativistic U+U collisions}},\
  }\href {https://doi.org/10.1140/epja/s10050-023-00982-0} {\bibfield
  {journal} {\bibinfo  {journal} {Eur. Phys. J. A}\ }\textbf {\bibinfo {volume}
  {59}},\ \bibinfo {pages} {64} (\bibinfo {year} {2023})}\BibitemShut {NoStop}%
\bibitem [{\citenamefont {{Stratonovich}}(1957)}]{Stratonovich1957}%
  \BibitemOpen
  \bibfield  {author} {\bibinfo {author} {\bibfnamefont {R.~L.}\ \bibnamefont
  {{Stratonovich}}},\ }\bibfield  {title} {\bibinfo {title} {{On a Method of
  Calculating Quantum Distribution Functions}},\ }\href@noop {} {\bibfield
  {journal} {\bibinfo  {journal} {Soviet Physics Doklady}\ }\textbf {\bibinfo
  {volume} {2}},\ \bibinfo {pages} {416} (\bibinfo {year} {1957})}\BibitemShut
  {NoStop}%
\bibitem [{\citenamefont {Hubbard}(1959)}]{Hubbard1959}%
  \BibitemOpen
  \bibfield  {author} {\bibinfo {author} {\bibfnamefont {J.}~\bibnamefont
  {Hubbard}},\ }\bibfield  {title} {\bibinfo {title} {{Calculation of partition
  functions}},\ }\href {https://doi.org/10.1103/PhysRevLett.3.77} {\bibfield
  {journal} {\bibinfo  {journal} {Phys. Rev. Lett.}\ }\textbf {\bibinfo
  {volume} {3}},\ \bibinfo {pages} {77} (\bibinfo {year} {1959})}\BibitemShut
  {NoStop}%
\bibitem [{\citenamefont {Ormand}\ \emph {et~al.}(1994)\citenamefont {Ormand},
  \citenamefont {Dean}, \citenamefont {Johnson}, \citenamefont {Lang},\ and\
  \citenamefont {Koonin}}]{Ormand1994}%
  \BibitemOpen
  \bibfield  {author} {\bibinfo {author} {\bibfnamefont {W.~E.}\ \bibnamefont
  {Ormand}}, \bibinfo {author} {\bibfnamefont {D.~J.}\ \bibnamefont {Dean}},
  \bibinfo {author} {\bibfnamefont {C.~W.}\ \bibnamefont {Johnson}}, \bibinfo
  {author} {\bibfnamefont {G.~H.}\ \bibnamefont {Lang}},\ and\ \bibinfo
  {author} {\bibfnamefont {S.~E.}\ \bibnamefont {Koonin}},\ }\bibfield  {title}
  {\bibinfo {title} {{Demonstration of the auxiliary-field Monte Carlo approach
  for sd-shell nuclei}},\ }\href {https://doi.org/10.1103/PhysRevC.49.1422}
  {\bibfield  {journal} {\bibinfo  {journal} {Phys. Rev. C}\ }\textbf {\bibinfo
  {volume} {49}},\ \bibinfo {pages} {1422} (\bibinfo {year}
  {1994})}\BibitemShut {NoStop}%
\bibitem [{\citenamefont {Alhassid}\ \emph {et~al.}(1999)\citenamefont
  {Alhassid}, \citenamefont {Liu},\ and\ \citenamefont
  {Nakada}}]{Alhassid1999}%
  \BibitemOpen
  \bibfield  {author} {\bibinfo {author} {\bibfnamefont {Y.}~\bibnamefont
  {Alhassid}}, \bibinfo {author} {\bibfnamefont {S.}~\bibnamefont {Liu}},\ and\
  \bibinfo {author} {\bibfnamefont {H.}~\bibnamefont {Nakada}},\ }\bibfield
  {title} {\bibinfo {title} {Particle-number reprojection in the shell model
  monte carlo method: Application to nuclear level densities},\ }\href
  {https://doi.org/10.1103/PhysRevLett.83.4265} {\bibfield  {journal} {\bibinfo
   {journal} {Phys. Rev. Lett.}\ }\textbf {\bibinfo {volume} {83}},\ \bibinfo
  {pages} {4265} (\bibinfo {year} {1999})}\BibitemShut {NoStop}%
\bibitem [{\citenamefont {Alhassid}\ \emph {et~al.}(2024)\citenamefont
  {Alhassid}, \citenamefont {Fanto},\ and\ \citenamefont
  {\"Ozen}}]{Alhassid2024}%
  \BibitemOpen
  \bibfield  {author} {\bibinfo {author} {\bibfnamefont {Y.}~\bibnamefont
  {Alhassid}}, \bibinfo {author} {\bibfnamefont {P.}~\bibnamefont {Fanto}},\
  and\ \bibinfo {author} {\bibfnamefont {C.}~\bibnamefont {\"Ozen}},\
  }\bibfield  {title} {\bibinfo {title} {{Circumventing the odd-particle-number
  sign problem in the shell model Monte Carlo method}},\ }\href
  {https://doi.org/10.1103/PhysRevC.110.L061303} {\bibfield  {journal}
  {\bibinfo  {journal} {Phys. Rev. C}\ }\textbf {\bibinfo {volume} {110}},\
  \bibinfo {pages} {L061303} (\bibinfo {year} {2024})}\BibitemShut {NoStop}%
\bibitem [{sup()}]{suppmat}%
  \BibitemOpen
  \href@noop {} {}\bibinfo {note} {See the Supplemental Material accompanying
  this article.}\BibitemShut {Stop}%
\bibitem [{Nak()}]{Nakada}%
  \BibitemOpen
  \href@noop {} {}\bibinfo {note} {{H. Nakada, private
  communication}}\BibitemShut {NoStop}%
\bibitem [{\citenamefont {Dufour}\ and\ \citenamefont
  {Zuker}(1996)}]{Dufour1996}%
  \BibitemOpen
  \bibfield  {author} {\bibinfo {author} {\bibfnamefont {M.}~\bibnamefont
  {Dufour}}\ and\ \bibinfo {author} {\bibfnamefont {A.}~\bibnamefont {Zuker}},\
  }\bibfield  {title} {\bibinfo {title} {{The realistic collective nuclear
  Hamiltonian}},\ }\href {https://doi.org/10.1103/PhysRevC.54.1641} {\bibfield
  {journal} {\bibinfo  {journal} {Phys. Rev. C}\ }\textbf {\bibinfo {volume}
  {54}},\ \bibinfo {pages} {1641} (\bibinfo {year} {1996})}\BibitemShut
  {NoStop}%
\bibitem [{\citenamefont {Alhassid}\ \emph {et~al.}(1996)\citenamefont
  {Alhassid}, \citenamefont {Bertsch}, \citenamefont {Dean},\ and\
  \citenamefont {Koonin}}]{Alhassid1996}%
  \BibitemOpen
  \bibfield  {author} {\bibinfo {author} {\bibfnamefont {Y.}~\bibnamefont
  {Alhassid}}, \bibinfo {author} {\bibfnamefont {G.~F.}\ \bibnamefont
  {Bertsch}}, \bibinfo {author} {\bibfnamefont {D.~J.}\ \bibnamefont {Dean}},\
  and\ \bibinfo {author} {\bibfnamefont {S.~E.}\ \bibnamefont {Koonin}},\
  }\bibfield  {title} {\bibinfo {title} {{Shell model Monte Carlo studies of
  gamma soft nuclei}},\ }\href {https://doi.org/10.1103/PhysRevLett.77.1444}
  {\bibfield  {journal} {\bibinfo  {journal} {Phys. Rev. Lett.}\ }\textbf
  {\bibinfo {volume} {77}},\ \bibinfo {pages} {1444} (\bibinfo {year}
  {1996})}\BibitemShut {NoStop}%
\bibitem [{\citenamefont {Gilbert}\ and\ \citenamefont
  {Cameron}(1965)}]{Gilbert1965}%
  \BibitemOpen
  \bibfield  {author} {\bibinfo {author} {\bibfnamefont {A.}~\bibnamefont
  {Gilbert}}\ and\ \bibinfo {author} {\bibfnamefont {A.~G.~W.}\ \bibnamefont
  {Cameron}},\ }\bibfield  {title} {\bibinfo {title} {{A composite
  nuclear-level density formula with shell corrections}},\ }\href
  {https://doi.org/10.1139/p65-139} {\bibfield  {journal} {\bibinfo  {journal}
  {Can. J. Phys.}\ }\textbf {\bibinfo {volume} {43}},\ \bibinfo {pages} {1446}
  (\bibinfo {year} {1965})}\BibitemShut {NoStop}%
\bibitem [{\citenamefont {Krieger}\ \emph {et~al.}(1990)\citenamefont
  {Krieger}, \citenamefont {Bonche}, \citenamefont {Flocard}, \citenamefont
  {Quentin},\ and\ \citenamefont {Weiss}}]{Krieger1990}%
  \BibitemOpen
  \bibfield  {author} {\bibinfo {author} {\bibfnamefont {S.~J.}\ \bibnamefont
  {Krieger}}, \bibinfo {author} {\bibfnamefont {P.}~\bibnamefont {Bonche}},
  \bibinfo {author} {\bibfnamefont {H.}~\bibnamefont {Flocard}}, \bibinfo
  {author} {\bibfnamefont {P.}~\bibnamefont {Quentin}},\ and\ \bibinfo {author}
  {\bibfnamefont {M.~S.}\ \bibnamefont {Weiss}},\ }\bibfield  {title} {\bibinfo
  {title} {{An improved pairing interaction for mean field calculations using
  skyrme potentials*}},\ }\href {https://doi.org/10.1016/0375-9474(90)90035-K}
  {\bibfield  {journal} {\bibinfo  {journal} {Nucl. Phys. A}\ }\textbf
  {\bibinfo {volume} {517}},\ \bibinfo {pages} {275} (\bibinfo {year}
  {1990})}\BibitemShut {NoStop}%
\bibitem [{\citenamefont {Cwiok}\ \emph {et~al.}(1996)\citenamefont {Cwiok},
  \citenamefont {Dobaczewski}, \citenamefont {Heenen}, \citenamefont
  {Magierski},\ and\ \citenamefont {Nazarewicz}}]{Cwiok1996}%
  \BibitemOpen
  \bibfield  {author} {\bibinfo {author} {\bibfnamefont {S.}~\bibnamefont
  {Cwiok}}, \bibinfo {author} {\bibfnamefont {J.}~\bibnamefont {Dobaczewski}},
  \bibinfo {author} {\bibfnamefont {P.~H.}\ \bibnamefont {Heenen}}, \bibinfo
  {author} {\bibfnamefont {P.}~\bibnamefont {Magierski}},\ and\ \bibinfo
  {author} {\bibfnamefont {W.}~\bibnamefont {Nazarewicz}},\ }\bibfield  {title}
  {\bibinfo {title} {{Shell structure of the superheavy elements}},\ }\href
  {https://doi.org/10.1016/S0375-9474(96)00337-5} {\bibfield  {journal}
  {\bibinfo  {journal} {Nucl. Phys. A}\ }\textbf {\bibinfo {volume} {611}},\
  \bibinfo {pages} {211} (\bibinfo {year} {1996})}\BibitemShut {NoStop}%
\bibitem [{\citenamefont {Donati}\ \emph {et~al.}(2005)\citenamefont {Donati},
  \citenamefont {Cori}, \citenamefont {Barranco}, \citenamefont {Broglia},\
  and\ \citenamefont {Vigezzi}}]{Donati2005}%
  \BibitemOpen
  \bibfield  {author} {\bibinfo {author} {\bibfnamefont {P.}~\bibnamefont
  {Donati}}, \bibinfo {author} {\bibfnamefont {G.}~\bibnamefont {Cori}},
  \bibinfo {author} {\bibfnamefont {F.}~\bibnamefont {Barranco}}, \bibinfo
  {author} {\bibfnamefont {R.}~\bibnamefont {Broglia}},\ and\ \bibinfo {author}
  {\bibfnamefont {E.}~\bibnamefont {Vigezzi}},\ }\bibfield  {title} {\bibinfo
  {title} {{Effective pairing interaction induced by polarization effects in
  deformed nuclei}},\ }\href {https://doi.org/10.1088/0954-3899/31/5/002}
  {\bibfield  {journal} {\bibinfo  {journal} {J. Phys. G: Nucl. Part. Phys.}\
  }\textbf {\bibinfo {volume} {31}},\ \bibinfo {pages} {295} (\bibinfo {year}
  {2005})}\BibitemShut {NoStop}%
\bibitem [{\citenamefont {Goriely}\ \emph {et~al.}(2008)\citenamefont
  {Goriely}, \citenamefont {Hilaire},\ and\ \citenamefont
  {Koning}}]{Goriely2008}%
  \BibitemOpen
  \bibfield  {author} {\bibinfo {author} {\bibfnamefont {S.}~\bibnamefont
  {Goriely}}, \bibinfo {author} {\bibfnamefont {S.}~\bibnamefont {Hilaire}},\
  and\ \bibinfo {author} {\bibfnamefont {A.~J.}\ \bibnamefont {Koning}},\
  }\bibfield  {title} {\bibinfo {title} {{Improved microscopic nuclear level
  densities within the Hartree-Fock-Bogoliubov plus combinatorial method}},\
  }\href {https://doi.org/10.1103/PhysRevC.78.064307} {\bibfield  {journal}
  {\bibinfo  {journal} {Phys. Rev. C}\ }\textbf {\bibinfo {volume} {78}},\
  \bibinfo {pages} {064307} (\bibinfo {year} {2008})}\BibitemShut {NoStop}%
\bibitem [{\citenamefont {Bonett-Matiz}\ \emph {et~al.}(2013)\citenamefont
  {Bonett-Matiz}, \citenamefont {Mukherjee},\ and\ \citenamefont
  {Alhassid}}]{Bonett-Matiz2013}%
  \BibitemOpen
  \bibfield  {author} {\bibinfo {author} {\bibfnamefont {M.}~\bibnamefont
  {Bonett-Matiz}}, \bibinfo {author} {\bibfnamefont {A.}~\bibnamefont
  {Mukherjee}},\ and\ \bibinfo {author} {\bibfnamefont {Y.}~\bibnamefont
  {Alhassid}},\ }\bibfield  {title} {\bibinfo {title} {{Level densities of
  nickel isotopes: microscopic theory versus experiment}},\ }\href
  {https://doi.org/10.1103/PhysRevC.88.011302} {\bibfield  {journal} {\bibinfo
  {journal} {Phys. Rev. C}\ }\textbf {\bibinfo {volume} {88}},\ \bibinfo
  {pages} {011302} (\bibinfo {year} {2013})}\BibitemShut {NoStop}%
\bibitem [{\citenamefont {Liu}\ and\ \citenamefont {Alhassid}(2001)}]{Liu2001}%
  \BibitemOpen
  \bibfield  {author} {\bibinfo {author} {\bibfnamefont {S.}~\bibnamefont
  {Liu}}\ and\ \bibinfo {author} {\bibfnamefont {Y.}~\bibnamefont {Alhassid}},\
  }\bibfield  {title} {\bibinfo {title} {{Signature of a pairing transition in
  the heat capacity of finite nuclei}},\ }\href
  {https://doi.org/10.1103/PhysRevLett.87.022501} {\bibfield  {journal}
  {\bibinfo  {journal} {Phys. Rev. Lett.}\ }\textbf {\bibinfo {volume} {87}},\
  \bibinfo {pages} {022501} (\bibinfo {year} {2001})}\BibitemShut {NoStop}%
\bibitem [{\citenamefont {\"Ozen}\ and\ \citenamefont
  {Alhassid}(2025)}]{Ozen2025}%
  \BibitemOpen
  \bibfield  {author} {\bibinfo {author} {\bibfnamefont {C.}~\bibnamefont
  {\"Ozen}}\ and\ \bibinfo {author} {\bibfnamefont {Y.}~\bibnamefont
  {Alhassid}},\ }\bibfield  {title} {\bibinfo {title} {{Direct local
  parametrization of nuclear state densities using the back-shifted Bethe
  formula}},\ }\href {https://doi.org/10.1016/j.nuclphysa.2025.123034}
  {\bibfield  {journal} {\bibinfo  {journal} {Nucl. Phys. A}\ }\textbf
  {\bibinfo {volume} {1058}},\ \bibinfo {pages} {123034} (\bibinfo {year}
  {2025})}\BibitemShut {NoStop}%
\bibitem [{\citenamefont {Capote}\ \emph {et~al.}(2009)\citenamefont {Capote},
  \citenamefont {Herman}, \citenamefont {Oblo\v{z}insk\'y}, \citenamefont
  {Young}, \citenamefont {Goriely}, \citenamefont {Belgya}, \citenamefont
  {Ignatyuk}, \citenamefont {Koning}, \citenamefont {Hilaire}, \citenamefont
  {Plujko} \emph {et~al.}}]{Capote2009}%
  \BibitemOpen
  \bibfield  {author} {\bibinfo {author} {\bibfnamefont {R.}~\bibnamefont
  {Capote}}, \bibinfo {author} {\bibfnamefont {M.}~\bibnamefont {Herman}},
  \bibinfo {author} {\bibfnamefont {P.}~\bibnamefont {Oblo\v{z}insk\'y}},
  \bibinfo {author} {\bibfnamefont {P.~G.}\ \bibnamefont {Young}}, \bibinfo
  {author} {\bibfnamefont {S.}~\bibnamefont {Goriely}}, \bibinfo {author}
  {\bibfnamefont {T.}~\bibnamefont {Belgya}}, \bibinfo {author} {\bibfnamefont
  {A.~V.}\ \bibnamefont {Ignatyuk}}, \bibinfo {author} {\bibfnamefont {A.~J.}\
  \bibnamefont {Koning}}, \bibinfo {author} {\bibfnamefont {S.}~\bibnamefont
  {Hilaire}}, \bibinfo {author} {\bibfnamefont {V.~A.}\ \bibnamefont {Plujko}},
  \emph {et~al.},\ }\bibfield  {title} {\bibinfo {title} {{RIPL – Reference
  Input Parameter Library for Calculation of Nuclear Reactions and Nuclear Data
  Evaluations}},\ }\href
  {https://doi.org/https://doi.org/10.1016/j.nds.2009.10.004} {\bibfield
  {journal} {\bibinfo  {journal} {Nuclear Data Sheets}\ }\textbf {\bibinfo
  {volume} {110}},\ \bibinfo {pages} {3107} (\bibinfo {year}
  {2009})}\BibitemShut {NoStop}%
\bibitem [{\citenamefont {Dilg}\ \emph {et~al.}(1973)\citenamefont {Dilg},
  \citenamefont {Schantl}, \citenamefont {Vonach},\ and\ \citenamefont
  {Uhl}}]{Dilg1973}%
  \BibitemOpen
  \bibfield  {author} {\bibinfo {author} {\bibfnamefont {W.}~\bibnamefont
  {Dilg}}, \bibinfo {author} {\bibfnamefont {W.}~\bibnamefont {Schantl}},
  \bibinfo {author} {\bibfnamefont {H.}~\bibnamefont {Vonach}},\ and\ \bibinfo
  {author} {\bibfnamefont {M.}~\bibnamefont {Uhl}},\ }\bibfield  {title}
  {\bibinfo {title} {{Level density parameters for the back-shifted fermi gas
  model in the mass range 4 0 \ensuremath{<} A \ensuremath{<} 250}},\ }\href
  {https://doi.org/10.1016/0375-9474(73)90196-6} {\bibfield  {journal}
  {\bibinfo  {journal} {Nucl. Phys. A}\ }\textbf {\bibinfo {volume} {217}},\
  \bibinfo {pages} {269} (\bibinfo {year} {1973})}\BibitemShut {NoStop}%
\bibitem [{\citenamefont {Goodman}(1981)}]{Goodman1981}%
  \BibitemOpen
  \bibfield  {author} {\bibinfo {author} {\bibfnamefont {A.~L.}\ \bibnamefont
  {Goodman}},\ }\bibfield  {title} {\bibinfo {title} {{Finite-temperature HFB
  theory}},\ }\href {https://doi.org/10.1016/0375-9474(81)90557-1} {\bibfield
  {journal} {\bibinfo  {journal} {Nucl. Phys. A}\ }\textbf {\bibinfo {volume}
  {352}},\ \bibinfo {pages} {30} (\bibinfo {year} {1981})}\BibitemShut
  {NoStop}%
\bibitem [{\citenamefont {Ryssens}\ and\ \citenamefont
  {Alhassid}(2021)}]{Ryssens2021}%
  \BibitemOpen
  \bibfield  {author} {\bibinfo {author} {\bibfnamefont {W.}~\bibnamefont
  {Ryssens}}\ and\ \bibinfo {author} {\bibfnamefont {Y.}~\bibnamefont
  {Alhassid}},\ }\bibfield  {title} {\bibinfo {title} {{Finite-temperature
  mean-field approximations for shell model Hamiltonians: the code HF-SHELL}},\
  }\href {https://doi.org/10.1140/epja/s10050-021-00365-3} {\bibfield
  {journal} {\bibinfo  {journal} {Eur. Phys. J. A}\ }\textbf {\bibinfo {volume}
  {57}},\ \bibinfo {pages} {76} (\bibinfo {year} {2021})}\BibitemShut {NoStop}%
\bibitem [{\citenamefont {Alhassid}\ \emph {et~al.}(2007)\citenamefont
  {Alhassid}, \citenamefont {Liu},\ and\ \citenamefont
  {Nakada}}]{Alhassid2007}%
  \BibitemOpen
  \bibfield  {author} {\bibinfo {author} {\bibfnamefont {Y.}~\bibnamefont
  {Alhassid}}, \bibinfo {author} {\bibfnamefont {S.}~\bibnamefont {Liu}},\ and\
  \bibinfo {author} {\bibfnamefont {H.}~\bibnamefont {Nakada}},\ }\bibfield
  {title} {\bibinfo {title} {{Spin projection in the shell model Monte Carlo
  method and the spin distribution of nuclear level densities}},\ }\href
  {https://doi.org/10.1103/PhysRevLett.99.162504} {\bibfield  {journal}
  {\bibinfo  {journal} {Phys. Rev. Lett.}\ }\textbf {\bibinfo {volume} {99}},\
  \bibinfo {pages} {162504} (\bibinfo {year} {2007})}\BibitemShut {NoStop}%
\bibitem [{\citenamefont {Alhassid}\ \emph {et~al.}(2015)\citenamefont
  {Alhassid}, \citenamefont {Bonett-Matiz}, \citenamefont {Liu},\ and\
  \citenamefont {Nakada}}]{Alhassid2015}%
  \BibitemOpen
  \bibfield  {author} {\bibinfo {author} {\bibfnamefont {Y.}~\bibnamefont
  {Alhassid}}, \bibinfo {author} {\bibfnamefont {M.}~\bibnamefont
  {Bonett-Matiz}}, \bibinfo {author} {\bibfnamefont {S.}~\bibnamefont {Liu}},\
  and\ \bibinfo {author} {\bibfnamefont {H.}~\bibnamefont {Nakada}},\
  }\bibfield  {title} {\bibinfo {title} {Direct microscopic calculation of
  nuclear level densities in the shell model monte carlo approach},\ }\href
  {https://doi.org/10.1103/PhysRevC.92.024307} {\bibfield  {journal} {\bibinfo
  {journal} {Phys. Rev. C}\ }\textbf {\bibinfo {volume} {92}},\ \bibinfo
  {pages} {024307} (\bibinfo {year} {2015})}\BibitemShut {NoStop}%
\bibitem [{\citenamefont {Rauscher}\ \emph {et~al.}(1997)\citenamefont
  {Rauscher}, \citenamefont {Thielemann},\ and\ \citenamefont
  {Kratz}}]{Rauscher1997}%
  \BibitemOpen
  \bibfield  {author} {\bibinfo {author} {\bibfnamefont {T.}~\bibnamefont
  {Rauscher}}, \bibinfo {author} {\bibfnamefont {F.-K.}\ \bibnamefont
  {Thielemann}},\ and\ \bibinfo {author} {\bibfnamefont {K.-L.}\ \bibnamefont
  {Kratz}},\ }\bibfield  {title} {\bibinfo {title} {{Nuclear level density and
  the determination of thermonuclear rates for astrophysics}},\ }\href
  {https://doi.org/10.1103/PhysRevC.56.1613} {\bibfield  {journal} {\bibinfo
  {journal} {Phys. Rev. C}\ }\textbf {\bibinfo {volume} {56}},\ \bibinfo
  {pages} {1613} (\bibinfo {year} {1997})}\BibitemShut {NoStop}%
\bibitem [{\citenamefont {Ericson}(1960)}]{Ericson1960}%
  \BibitemOpen
  \bibfield  {author} {\bibinfo {author} {\bibfnamefont {T.}~\bibnamefont
  {Ericson}},\ }\bibfield  {title} {\bibinfo {title} {The statistical model and
  nuclear level densities},\ }\href {https://doi.org/10.1080/00018736000101239}
  {\bibfield  {journal} {\bibinfo  {journal} {Advances in Physics}\ }\textbf
  {\bibinfo {volume} {9}},\ \bibinfo {pages} {425} (\bibinfo {year}
  {1960})}\BibitemShut {NoStop}%
\bibitem [{\citenamefont {Zhongfu}\ \emph {et~al.}(1991)\citenamefont
  {Zhongfu}, \citenamefont {Ping}, \citenamefont {Zongdi},\ and\ \citenamefont
  {Chunmei}}]{Zhongfu1991}%
  \BibitemOpen
  \bibfield  {author} {\bibinfo {author} {\bibfnamefont {H.}~\bibnamefont
  {Zhongfu}}, \bibinfo {author} {\bibfnamefont {H.}~\bibnamefont {Ping}},
  \bibinfo {author} {\bibfnamefont {S.}~\bibnamefont {Zongdi}},\ and\ \bibinfo
  {author} {\bibfnamefont {Z.}~\bibnamefont {Chunmei}},\ }\href@noop {}
  {\bibfield  {journal} {\bibinfo  {journal} {Chin. J. Nucl. Phys.}\ }\textbf
  {\bibinfo {volume} {13}},\ \bibinfo {pages} {147} (\bibinfo {year}
  {1991})}\BibitemShut {NoStop}%
\bibitem [{\citenamefont {von Egidy}\ and\ \citenamefont
  {Bucurescu}(2005)}]{vonEgidy2005}%
  \BibitemOpen
  \bibfield  {author} {\bibinfo {author} {\bibfnamefont {T.}~\bibnamefont {von
  Egidy}}\ and\ \bibinfo {author} {\bibfnamefont {D.}~\bibnamefont
  {Bucurescu}},\ }\bibfield  {title} {\bibinfo {title} {{Systematics of nuclear
  level density parameters}},\ }\href
  {https://doi.org/10.1103/PhysRevC.72.044311} {\bibfield  {journal} {\bibinfo
  {journal} {Phys. Rev. C}\ }\textbf {\bibinfo {volume} {72}},\ \bibinfo
  {pages} {044311} (\bibinfo {year} {2005})},\ \bibinfo {note} {[Erratum:
  Phys.Rev.C 73, 049901 (2006)]}\BibitemShut {NoStop}%
\bibitem [{dat()}]{data}%
  \BibitemOpen
  \href@noop {} {}\bibinfo {note}
  {\lowercase{h}ttps://doi.org/10.5281/zenodo.18602660}\BibitemShut {NoStop}%
\end{thebibliography}%


\begin{thebibliography}{17}%
\makeatletter
\providecommand \@ifxundefined [1]{%
 \@ifx{#1\undefined}
}%
\providecommand \@ifnum [1]{%
 \ifnum #1\expandafter \@firstoftwo
 \else \expandafter \@secondoftwo
 \fi
}%
\providecommand \@ifx [1]{%
 \ifx #1\expandafter \@firstoftwo
 \else \expandafter \@secondoftwo
 \fi
}%
\providecommand \natexlab [1]{#1}%
\providecommand \enquote  [1]{``#1''}%
\providecommand \bibnamefont  [1]{#1}%
\providecommand \bibfnamefont [1]{#1}%
\providecommand \citenamefont [1]{#1}%
\providecommand \href@noop [0]{\@secondoftwo}%
\providecommand \href [0]{\begingroup \@sanitize@url \@href}%
\providecommand \@href[1]{\@@startlink{#1}\@@href}%
\providecommand \@@href[1]{\endgroup#1\@@endlink}%
\providecommand \@sanitize@url [0]{\catcode `\\12\catcode `\$12\catcode
  `\&12\catcode `\#12\catcode `\^12\catcode `\_12\catcode `\%12\relax}%
\providecommand \@@startlink[1]{}%
\providecommand \@@endlink[0]{}%
\providecommand \url  [0]{\begingroup\@sanitize@url \@url }%
\providecommand \@url [1]{\endgroup\@href {#1}{\urlprefix }}%
\providecommand \urlprefix  [0]{URL }%
\providecommand \Eprint [0]{\href }%
\providecommand \doibase [0]{https://doi.org/}%
\providecommand \selectlanguage [0]{\@gobble}%
\providecommand \bibinfo  [0]{\@secondoftwo}%
\providecommand \bibfield  [0]{\@secondoftwo}%
\providecommand \translation [1]{[#1]}%
\providecommand \BibitemOpen [0]{}%
\providecommand \bibitemStop [0]{}%
\providecommand \bibitemNoStop [0]{.\EOS\space}%
\providecommand \EOS [0]{\spacefactor3000\relax}%
\providecommand \BibitemShut  [1]{\csname bibitem#1\endcsname}%
\let\auto@bib@innerbib\@empty
\bibitem [{\citenamefont {\"Ozen}\ \emph {et~al.}(2015)\citenamefont {\"Ozen},
  \citenamefont {Alhassid},\ and\ \citenamefont {Nakada}}]{Ozen2015}%
  \BibitemOpen
  \bibfield  {author} {\bibinfo {author} {\bibfnamefont {C.}~\bibnamefont
  {\"Ozen}}, \bibinfo {author} {\bibfnamefont {Y.}~\bibnamefont {Alhassid}},\
  and\ \bibinfo {author} {\bibfnamefont {H.}~\bibnamefont {Nakada}},\
  }\bibfield  {title} {\bibinfo {title} {{Nuclear state densities of odd-mass
  heavy nuclei in the shell model Monte Carlo approach}},\ }\href
  {https://doi.org/10.1103/PhysRevC.91.034329} {\bibfield  {journal} {\bibinfo
  {journal} {Phys. Rev. C}\ }\textbf {\bibinfo {volume} {91}},\ \bibinfo
  {pages} {034329} (\bibinfo {year} {2015})}\BibitemShut {NoStop}%
\bibitem [{\citenamefont {Alhassid}\ \emph {et~al.}(2024)\citenamefont
  {Alhassid}, \citenamefont {Fanto},\ and\ \citenamefont
  {\"Ozen}}]{Alhassid2024}%
  \BibitemOpen
  \bibfield  {author} {\bibinfo {author} {\bibfnamefont {Y.}~\bibnamefont
  {Alhassid}}, \bibinfo {author} {\bibfnamefont {P.}~\bibnamefont {Fanto}},\
  and\ \bibinfo {author} {\bibfnamefont {C.}~\bibnamefont {\"Ozen}},\
  }\bibfield  {title} {\bibinfo {title} {{Circumventing the odd-particle-number
  sign problem in the shell model Monte Carlo method}},\ }\href
  {https://doi.org/10.1103/PhysRevC.110.L061303} {\bibfield  {journal}
  {\bibinfo  {journal} {Phys. Rev. C}\ }\textbf {\bibinfo {volume} {110}},\
  \bibinfo {pages} {L061303} (\bibinfo {year} {2024})}\BibitemShut {NoStop}%
\bibitem [{\citenamefont {Guttormsen}\ \emph {et~al.}(2013)\citenamefont
  {Guttormsen}, \citenamefont {Jurado}, \citenamefont {Wilson}, \citenamefont
  {Aiche}, \citenamefont {Bernstein}, \citenamefont {Ducasse}, \citenamefont
  {Giacoppo}, \citenamefont {Gorgen}, \citenamefont {Gunsing}, \citenamefont
  {Hagen} \emph {et~al.}}]{Guttormsen2013}%
  \BibitemOpen
  \bibfield  {author} {\bibinfo {author} {\bibfnamefont {M.}~\bibnamefont
  {Guttormsen}}, \bibinfo {author} {\bibfnamefont {B.}~\bibnamefont {Jurado}},
  \bibinfo {author} {\bibfnamefont {J.~N.}\ \bibnamefont {Wilson}}, \bibinfo
  {author} {\bibfnamefont {M.}~\bibnamefont {Aiche}}, \bibinfo {author}
  {\bibfnamefont {L.~A.}\ \bibnamefont {Bernstein}}, \bibinfo {author}
  {\bibfnamefont {Q.}~\bibnamefont {Ducasse}}, \bibinfo {author} {\bibfnamefont
  {F.}~\bibnamefont {Giacoppo}}, \bibinfo {author} {\bibfnamefont
  {A.}~\bibnamefont {Gorgen}}, \bibinfo {author} {\bibfnamefont
  {F.}~\bibnamefont {Gunsing}}, \bibinfo {author} {\bibfnamefont {T.~W.}\
  \bibnamefont {Hagen}}, \emph {et~al.},\ }\bibfield  {title} {\bibinfo {title}
  {{Constant-temperature level densities in the quasicontinuum of Th and U
  isotopes}},\ }\href {https://doi.org/10.1103/PhysRevC.88.024307} {\bibfield
  {journal} {\bibinfo  {journal} {Phys. Rev. C}\ }\textbf {\bibinfo {volume}
  {88}},\ \bibinfo {pages} {024307} (\bibinfo {year} {2013})}\BibitemShut
  {NoStop}%
\bibitem [{\citenamefont {Laplace}\ \emph {et~al.}(2016)\citenamefont
  {Laplace}, \citenamefont {Zeiser}, \citenamefont {Guttormsen}, \citenamefont
  {Larsen}, \citenamefont {Bleuel}, \citenamefont {Bernstein}, \citenamefont
  {Goldblum}, \citenamefont {Siem}, \citenamefont {Bello~Garotte},
  \citenamefont {Brown} \emph {et~al.}}]{Laplace2016}%
  \BibitemOpen
  \bibfield  {author} {\bibinfo {author} {\bibfnamefont {T.~A.}\ \bibnamefont
  {Laplace}}, \bibinfo {author} {\bibfnamefont {F.}~\bibnamefont {Zeiser}},
  \bibinfo {author} {\bibfnamefont {M.}~\bibnamefont {Guttormsen}}, \bibinfo
  {author} {\bibfnamefont {A.~C.}\ \bibnamefont {Larsen}}, \bibinfo {author}
  {\bibfnamefont {D.~L.}\ \bibnamefont {Bleuel}}, \bibinfo {author}
  {\bibfnamefont {L.~A.}\ \bibnamefont {Bernstein}}, \bibinfo {author}
  {\bibfnamefont {B.~L.}\ \bibnamefont {Goldblum}}, \bibinfo {author}
  {\bibfnamefont {S.}~\bibnamefont {Siem}}, \bibinfo {author} {\bibfnamefont
  {F.~L.}\ \bibnamefont {Bello~Garotte}}, \bibinfo {author} {\bibfnamefont
  {J.~A.}\ \bibnamefont {Brown}}, \emph {et~al.},\ }\bibfield  {title}
  {\bibinfo {title} {Statistical properties of $^{243}\text{Pu}$, and
  $^{242}\text{Pu}(n,\ensuremath{\gamma})$ cross section calculation},\ }\href
  {https://doi.org/10.1103/PhysRevC.93.014323} {\bibfield  {journal} {\bibinfo
  {journal} {Phys. Rev. C}\ }\textbf {\bibinfo {volume} {93}},\ \bibinfo
  {pages} {014323} (\bibinfo {year} {2016})}\BibitemShut {NoStop}%
\bibitem [{\citenamefont {Zeiser}\ \emph {et~al.}(2019)\citenamefont {Zeiser},
  \citenamefont {Tveten}, \citenamefont {Potel}, \citenamefont {Larsen},
  \citenamefont {Guttormsen}, \citenamefont {Laplace}, \citenamefont {Siem},
  \citenamefont {Bleuel}, \citenamefont {Goldblum}, \citenamefont {Bernstein}
  \emph {et~al.}}]{Zeiser2019}%
  \BibitemOpen
  \bibfield  {author} {\bibinfo {author} {\bibfnamefont {F.}~\bibnamefont
  {Zeiser}}, \bibinfo {author} {\bibfnamefont {G.~M.}\ \bibnamefont {Tveten}},
  \bibinfo {author} {\bibfnamefont {G.}~\bibnamefont {Potel}}, \bibinfo
  {author} {\bibfnamefont {A.~C.}\ \bibnamefont {Larsen}}, \bibinfo {author}
  {\bibfnamefont {M.}~\bibnamefont {Guttormsen}}, \bibinfo {author}
  {\bibfnamefont {T.~A.}\ \bibnamefont {Laplace}}, \bibinfo {author}
  {\bibfnamefont {S.}~\bibnamefont {Siem}}, \bibinfo {author} {\bibfnamefont
  {D.~L.}\ \bibnamefont {Bleuel}}, \bibinfo {author} {\bibfnamefont {B.~L.}\
  \bibnamefont {Goldblum}}, \bibinfo {author} {\bibfnamefont {L.~A.}\
  \bibnamefont {Bernstein}}, \emph {et~al.},\ }\bibfield  {title} {\bibinfo
  {title} {Restricted spin-range correction in the oslo method: The example of
  nuclear level density and $\ensuremath{\gamma}$-ray strength function from
  $^{239}\mathrm{Pu}(d,p\ensuremath{\gamma})^{240}\mathrm{Pu}$},\ }\href
  {https://doi.org/10.1103/PhysRevC.100.024305} {\bibfield  {journal} {\bibinfo
   {journal} {Phys. Rev. C}\ }\textbf {\bibinfo {volume} {100}},\ \bibinfo
  {pages} {024305} (\bibinfo {year} {2019})}\BibitemShut {NoStop}%
\bibitem [{\citenamefont {von Egidy}\ and\ \citenamefont
  {Bucurescu}(2005)}]{vonEgidy2005}%
  \BibitemOpen
  \bibfield  {author} {\bibinfo {author} {\bibfnamefont {T.}~\bibnamefont {von
  Egidy}}\ and\ \bibinfo {author} {\bibfnamefont {D.}~\bibnamefont
  {Bucurescu}},\ }\bibfield  {title} {\bibinfo {title} {{Systematics of nuclear
  level density parameters}},\ }\href
  {https://doi.org/10.1103/PhysRevC.72.044311} {\bibfield  {journal} {\bibinfo
  {journal} {Phys. Rev. C}\ }\textbf {\bibinfo {volume} {72}},\ \bibinfo
  {pages} {044311} (\bibinfo {year} {2005})},\ \bibinfo {note} {[Erratum:
  Phys.Rev.C 73, 049901 (2006)]}\BibitemShut {NoStop}%
\bibitem [{\citenamefont {\"Ozen}\ and\ \citenamefont
  {Alhassid}(2025)}]{Ozen2025}%
  \BibitemOpen
  \bibfield  {author} {\bibinfo {author} {\bibfnamefont {C.}~\bibnamefont
  {\"Ozen}}\ and\ \bibinfo {author} {\bibfnamefont {Y.}~\bibnamefont
  {Alhassid}},\ }\bibfield  {title} {\bibinfo {title} {{Direct local
  parametrization of nuclear state densities using the back-shifted Bethe
  formula}},\ }\href {https://doi.org/10.1016/j.nuclphysa.2025.123034}
  {\bibfield  {journal} {\bibinfo  {journal} {Nucl. Phys. A}\ }\textbf
  {\bibinfo {volume} {1058}},\ \bibinfo {pages} {123034} (\bibinfo {year}
  {2025})}\BibitemShut {NoStop}%
\bibitem [{\citenamefont {Larsen}\ \emph {et~al.}(2011)\citenamefont {Larsen},
  \citenamefont {Guttormsen}, \citenamefont {Krticka}, \citenamefont {Betak},
  \citenamefont {Burger}, \citenamefont {Gorgen}, \citenamefont {Nyhus},
  \citenamefont {Rekstad}, \citenamefont {Schiller}, \citenamefont {Siem} \emph
  {et~al.}}]{Larsen2011}%
  \BibitemOpen
  \bibfield  {author} {\bibinfo {author} {\bibfnamefont {A.~C.}\ \bibnamefont
  {Larsen}}, \bibinfo {author} {\bibfnamefont {M.}~\bibnamefont {Guttormsen}},
  \bibinfo {author} {\bibfnamefont {M.}~\bibnamefont {Krticka}}, \bibinfo
  {author} {\bibfnamefont {E.}~\bibnamefont {Betak}}, \bibinfo {author}
  {\bibfnamefont {A.}~\bibnamefont {Burger}}, \bibinfo {author} {\bibfnamefont
  {A.}~\bibnamefont {Gorgen}}, \bibinfo {author} {\bibfnamefont {H.~T.}\
  \bibnamefont {Nyhus}}, \bibinfo {author} {\bibfnamefont {J.}~\bibnamefont
  {Rekstad}}, \bibinfo {author} {\bibfnamefont {A.}~\bibnamefont {Schiller}},
  \bibinfo {author} {\bibfnamefont {S.}~\bibnamefont {Siem}}, \emph {et~al.},\
  }\bibfield  {title} {\bibinfo {title} {{Analysis of possible systematic
  errors in the Oslo method}},\ }\href
  {https://doi.org/10.1103/PhysRevC.83.034315} {\bibfield  {journal} {\bibinfo
  {journal} {Phys. Rev. C}\ }\textbf {\bibinfo {volume} {83}},\ \bibinfo
  {pages} {034315} (\bibinfo {year} {2011})},\ \bibinfo {note} {[Erratum:
  Phys.Rev.C 97, 049901 (2018)]}\BibitemShut {NoStop}%
\bibitem [{\citenamefont {Gilbert}\ and\ \citenamefont
  {Cameron}(1965)}]{Gilbert1965}%
  \BibitemOpen
  \bibfield  {author} {\bibinfo {author} {\bibfnamefont {A.}~\bibnamefont
  {Gilbert}}\ and\ \bibinfo {author} {\bibfnamefont {A.~G.~W.}\ \bibnamefont
  {Cameron}},\ }\bibfield  {title} {\bibinfo {title} {{A composite
  nuclear-level density formula with shell corrections}},\ }\href
  {https://doi.org/10.1139/p65-139} {\bibfield  {journal} {\bibinfo  {journal}
  {Can. J. Phys.}\ }\textbf {\bibinfo {volume} {43}},\ \bibinfo {pages} {1446}
  (\bibinfo {year} {1965})}\BibitemShut {NoStop}%
\bibitem [{\citenamefont {Goriely}\ \emph {et~al.}(2008)\citenamefont
  {Goriely}, \citenamefont {Hilaire},\ and\ \citenamefont
  {Koning}}]{Goriely2008}%
  \BibitemOpen
  \bibfield  {author} {\bibinfo {author} {\bibfnamefont {S.}~\bibnamefont
  {Goriely}}, \bibinfo {author} {\bibfnamefont {S.}~\bibnamefont {Hilaire}},\
  and\ \bibinfo {author} {\bibfnamefont {A.~J.}\ \bibnamefont {Koning}},\
  }\bibfield  {title} {\bibinfo {title} {{Improved microscopic nuclear level
  densities within the Hartree-Fock-Bogoliubov plus combinatorial method}},\
  }\href {https://doi.org/10.1103/PhysRevC.78.064307} {\bibfield  {journal}
  {\bibinfo  {journal} {Phys. Rev. C}\ }\textbf {\bibinfo {volume} {78}},\
  \bibinfo {pages} {064307} (\bibinfo {year} {2008})}\BibitemShut {NoStop}%
\bibitem [{\citenamefont {Goodman}(1981)}]{Goodman1981}%
  \BibitemOpen
  \bibfield  {author} {\bibinfo {author} {\bibfnamefont {A.~L.}\ \bibnamefont
  {Goodman}},\ }\bibfield  {title} {\bibinfo {title} {{Finite-temperature HFB
  theory}},\ }\href {https://doi.org/10.1016/0375-9474(81)90557-1} {\bibfield
  {journal} {\bibinfo  {journal} {Nucl. Phys. A}\ }\textbf {\bibinfo {volume}
  {352}},\ \bibinfo {pages} {30} (\bibinfo {year} {1981})}\BibitemShut
  {NoStop}%
\bibitem [{\citenamefont {Ryssens}\ and\ \citenamefont
  {Alhassid}(2021)}]{Ryssens2021}%
  \BibitemOpen
  \bibfield  {author} {\bibinfo {author} {\bibfnamefont {W.}~\bibnamefont
  {Ryssens}}\ and\ \bibinfo {author} {\bibfnamefont {Y.}~\bibnamefont
  {Alhassid}},\ }\bibfield  {title} {\bibinfo {title} {{Finite-temperature
  mean-field approximations for shell model Hamiltonians: the code HF-SHELL}},\
  }\href {https://doi.org/10.1140/epja/s10050-021-00365-3} {\bibfield
  {journal} {\bibinfo  {journal} {Eur. Phys. J. A}\ }\textbf {\bibinfo {volume}
  {57}},\ \bibinfo {pages} {76} (\bibinfo {year} {2021})}\BibitemShut {NoStop}%
\bibitem [{\citenamefont {Krieger}\ \emph {et~al.}(1990)\citenamefont
  {Krieger}, \citenamefont {Bonche}, \citenamefont {Flocard}, \citenamefont
  {Quentin},\ and\ \citenamefont {Weiss}}]{Krieger1990}%
  \BibitemOpen
  \bibfield  {author} {\bibinfo {author} {\bibfnamefont {S.~J.}\ \bibnamefont
  {Krieger}}, \bibinfo {author} {\bibfnamefont {P.}~\bibnamefont {Bonche}},
  \bibinfo {author} {\bibfnamefont {H.}~\bibnamefont {Flocard}}, \bibinfo
  {author} {\bibfnamefont {P.}~\bibnamefont {Quentin}},\ and\ \bibinfo {author}
  {\bibfnamefont {M.~S.}\ \bibnamefont {Weiss}},\ }\bibfield  {title} {\bibinfo
  {title} {{An improved pairing interaction for mean field calculations using
  skyrme potentials*}},\ }\href {https://doi.org/10.1016/0375-9474(90)90035-K}
  {\bibfield  {journal} {\bibinfo  {journal} {Nucl. Phys. A}\ }\textbf
  {\bibinfo {volume} {517}},\ \bibinfo {pages} {275} (\bibinfo {year}
  {1990})}\BibitemShut {NoStop}%
\bibitem [{\citenamefont {Cwiok}\ \emph {et~al.}(1996)\citenamefont {Cwiok},
  \citenamefont {Dobaczewski}, \citenamefont {Heenen}, \citenamefont
  {Magierski},\ and\ \citenamefont {Nazarewicz}}]{Cwiok1996}%
  \BibitemOpen
  \bibfield  {author} {\bibinfo {author} {\bibfnamefont {S.}~\bibnamefont
  {Cwiok}}, \bibinfo {author} {\bibfnamefont {J.}~\bibnamefont {Dobaczewski}},
  \bibinfo {author} {\bibfnamefont {P.~H.}\ \bibnamefont {Heenen}}, \bibinfo
  {author} {\bibfnamefont {P.}~\bibnamefont {Magierski}},\ and\ \bibinfo
  {author} {\bibfnamefont {W.}~\bibnamefont {Nazarewicz}},\ }\bibfield  {title}
  {\bibinfo {title} {{Shell structure of the superheavy elements}},\ }\href
  {https://doi.org/10.1016/S0375-9474(96)00337-5} {\bibfield  {journal}
  {\bibinfo  {journal} {Nucl. Phys. A}\ }\textbf {\bibinfo {volume} {611}},\
  \bibinfo {pages} {211} (\bibinfo {year} {1996})}\BibitemShut {NoStop}%
\bibitem [{\citenamefont {Donati}\ \emph {et~al.}(2005)\citenamefont {Donati},
  \citenamefont {Cori}, \citenamefont {Barranco}, \citenamefont {Broglia},\
  and\ \citenamefont {Vigezzi}}]{Donati2005}%
  \BibitemOpen
  \bibfield  {author} {\bibinfo {author} {\bibfnamefont {P.}~\bibnamefont
  {Donati}}, \bibinfo {author} {\bibfnamefont {G.}~\bibnamefont {Cori}},
  \bibinfo {author} {\bibfnamefont {F.}~\bibnamefont {Barranco}}, \bibinfo
  {author} {\bibfnamefont {R.}~\bibnamefont {Broglia}},\ and\ \bibinfo {author}
  {\bibfnamefont {E.}~\bibnamefont {Vigezzi}},\ }\bibfield  {title} {\bibinfo
  {title} {{Effective pairing interaction induced by polarization effects in
  deformed nuclei}},\ }\href {https://doi.org/10.1088/0954-3899/31/5/002}
  {\bibfield  {journal} {\bibinfo  {journal} {J. Phys. G: Nucl. Part. Phys.}\
  }\textbf {\bibinfo {volume} {31}},\ \bibinfo {pages} {295} (\bibinfo {year}
  {2005})}\BibitemShut {NoStop}%
\bibitem [{\citenamefont {Capote}\ \emph {et~al.}(2009)\citenamefont {Capote},
  \citenamefont {Herman}, \citenamefont {Oblo\v{z}insk\'y}, \citenamefont
  {Young}, \citenamefont {Goriely}, \citenamefont {Belgya}, \citenamefont
  {Ignatyuk}, \citenamefont {Koning}, \citenamefont {Hilaire}, \citenamefont
  {Plujko} \emph {et~al.}}]{Capote2009}%
  \BibitemOpen
  \bibfield  {author} {\bibinfo {author} {\bibfnamefont {R.}~\bibnamefont
  {Capote}}, \bibinfo {author} {\bibfnamefont {M.}~\bibnamefont {Herman}},
  \bibinfo {author} {\bibfnamefont {P.}~\bibnamefont {Oblo\v{z}insk\'y}},
  \bibinfo {author} {\bibfnamefont {P.~G.}\ \bibnamefont {Young}}, \bibinfo
  {author} {\bibfnamefont {S.}~\bibnamefont {Goriely}}, \bibinfo {author}
  {\bibfnamefont {T.}~\bibnamefont {Belgya}}, \bibinfo {author} {\bibfnamefont
  {A.~V.}\ \bibnamefont {Ignatyuk}}, \bibinfo {author} {\bibfnamefont {A.~J.}\
  \bibnamefont {Koning}}, \bibinfo {author} {\bibfnamefont {S.}~\bibnamefont
  {Hilaire}}, \bibinfo {author} {\bibfnamefont {V.~A.}\ \bibnamefont {Plujko}},
  \emph {et~al.},\ }\bibfield  {title} {\bibinfo {title} {{RIPL – Reference
  Input Parameter Library for Calculation of Nuclear Reactions and Nuclear Data
  Evaluations}},\ }\href
  {https://doi.org/https://doi.org/10.1016/j.nds.2009.10.004} {\bibfield
  {journal} {\bibinfo  {journal} {Nuclear Data Sheets}\ }\textbf {\bibinfo
  {volume} {110}},\ \bibinfo {pages} {3107} (\bibinfo {year}
  {2009})}\BibitemShut {NoStop}%
\bibitem [{\citenamefont {Alhassid}\ \emph {et~al.}(2007)\citenamefont
  {Alhassid}, \citenamefont {Liu},\ and\ \citenamefont
  {Nakada}}]{Alhassid2007}%
  \BibitemOpen
  \bibfield  {author} {\bibinfo {author} {\bibfnamefont {Y.}~\bibnamefont
  {Alhassid}}, \bibinfo {author} {\bibfnamefont {S.}~\bibnamefont {Liu}},\ and\
  \bibinfo {author} {\bibfnamefont {H.}~\bibnamefont {Nakada}},\ }\bibfield
  {title} {\bibinfo {title} {{Spin projection in the shell model Monte Carlo
  method and the spin distribution of nuclear level densities}},\ }\href
  {https://doi.org/10.1103/PhysRevLett.99.162504} {\bibfield  {journal}
  {\bibinfo  {journal} {Phys. Rev. Lett.}\ }\textbf {\bibinfo {volume} {99}},\
  \bibinfo {pages} {162504} (\bibinfo {year} {2007})}\BibitemShut {NoStop}%
\end{thebibliography}%

\appendix

\section{End matter}

\ssec{Results for other actinides}
In the following we present results for nine other actinides for which Oslo method experiments have not been performed but for which neutron resonance data and low-lying level counting data exist. The NSDs, NLDs, and spin distributions at the respective neutron separation energies have been calculated for these nuclei using the same methods as those presented in the main text. 

\begin{widetext}
\begin{figure*}[bth]
	\includegraphics[width=0.95\linewidth]{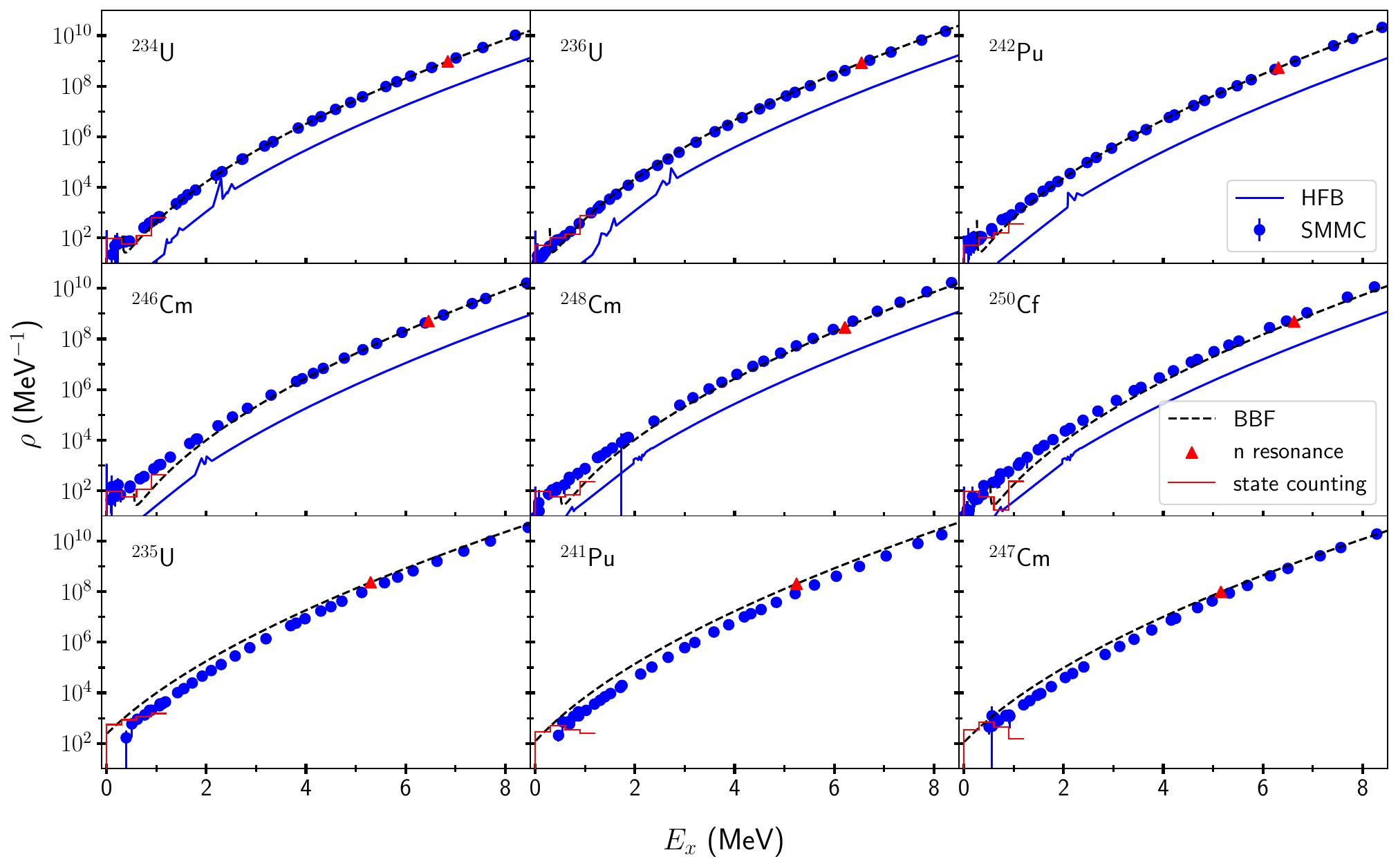}
	\caption{As in Fig.~\ref{fig_NSD}, but for other actinides.}
	\label{fig_NSD2}
\end{figure*}
\end{widetext}

\begin{figure*}[bth]
	\includegraphics[width=0.95\linewidth]{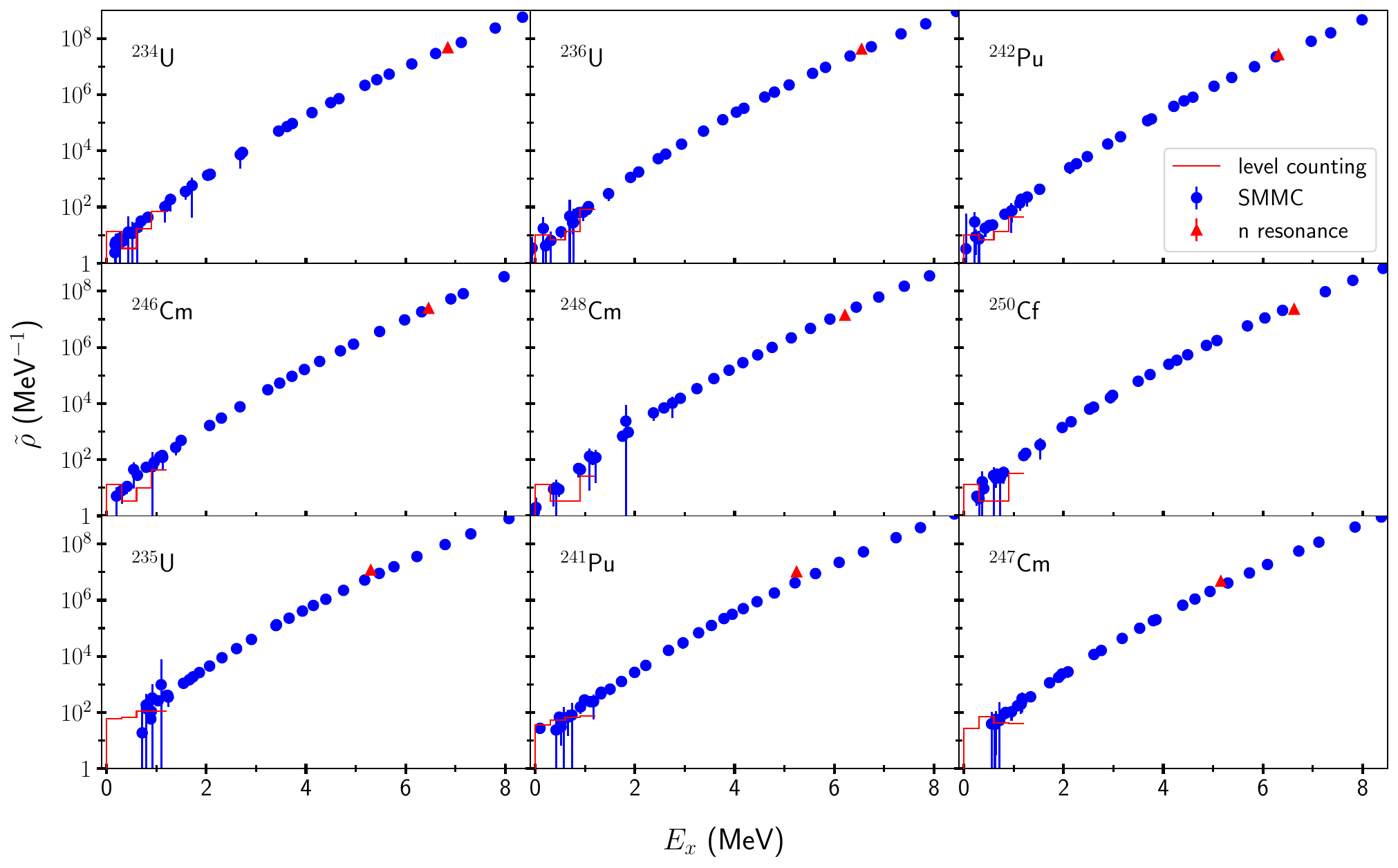}
	\caption{As in Fig.~\ref{fig_NLDs}, but for other actinides.}
	\label{fig_NLD2}
\end{figure*}

Figures \ref{fig_NSD2} and \ref{fig_NLD2} show the NSDs and NLDs, respectively, for these nuclei. The NSDs of these nuclei show similarly good agreement with the BBF densities as the actinides in the main text. Table \ref{table_sn2} provides the corresponding experimental and calculated values of the NLDs, spin-cutoff parameters and $D_0$ at the neutron separation energies.

\begin{table*}[b]
	\begin{tabular}{c|c|c|c|c|c|c|c|c}
		\hline
		\hline
		Nucleus & $S_n$ (MeV) & $J_t$ & \multicolumn{2}{|c|}{$\tilde{\rho}(S_n)$ ($10^6$ MeV$^{-1}$)} & \multicolumn{2}{c|}{$\sigma(S_n)$} & \multicolumn{2}{|c}{$D_0$ (eV)} \\
		& & & \multicolumn{2}{c|}{\hspace{0.2cm} Exp. \hspace{0.8cm} SMMC} & \multicolumn{2}{c|}{rigid body \hspace{0.4cm} SMMC} & \multicolumn{2}{c}{\hspace{0.2cm} Exp. \hspace{0.8cm} SMMC} \\
		\hline
		$^{234}$U & 6.845 & 5/2 & 46.8 $\pm$ 8.9 & 49.4 $\pm$ 5.6 &\phantom{rigid} 8.24 \phantom{rigid}& 8.64 $\pm$ 0.13 & 0.52 $\pm$ 0.02 & 0.54 $\pm$ 0.11\\
		$^{235}$U & 5.298 & 0 & 11.7 $\pm$ 2.5 & 6.74 $\pm$ 0.74 & 8.07 & 8.11 $\pm$ 0.15 & $11.2 \pm 0.8$ & 19.7 $\pm$ 2.3 \\
		$^{236}$U & 6.546 & 7/2 & 41.9 $\pm$ 7.8 & 38.3 $\pm$ 3.2& 8.15 & 8.56 $\pm$ 0.12 & 0.45 $\pm$ 0.03  & 0.53 $\pm$ 0.07 \\
		$^{241}$Pu & 5.242 & 0 & $10.2 \pm 2.1$ & 4.48 $\pm$ 0.57& 8.12 & 8.12 $\pm$ 0.18 & $13.0 \pm 0.5$ & 19.7 $\pm$ 4.0 \\
		$^{242}$Pu & 6.310 & 5/2 & 26.3 $\pm$ 5.4 & 25.9 $\pm$ 5.3& 8.27 & 8.61 $\pm$ 0.17  & 0.93 $\pm$ 0.08 & 1.02 $\pm$ 0.23 \\	
		$^{246}$Cm & 6.458 & 7/2 & 24.6 $\pm$ 4.6 & 26.7 $\pm$ 4.3& 8.30 & 8.60 $\pm$ 0.18 & 0.79 $\pm$ 0.05 & 0.78 $\pm$ 0.18 \\	
		$^{247}$Cm & 5.156 & 0 &  $4.71 \pm 1.22$  & 3.29 $\pm$ 0.42& 8.37 & 8.02 $\pm$ 0.20 & 30.0 $\pm$ 5.0 & 39.4 $\pm$ 5.4\\
		$^{248}$Cm & 6.213  & 9/2 & 13.9 $\pm$ 4.1 & 19.8 $\pm$ 2.4& 8.32 & 8.47 $\pm$ 0.18 & 1.2 $\pm$ 0.3 & 0.87 $\pm$ 0.11\\	
		$^{250}$Cf & 6.624  & 9/2 & 22.7 $\pm$ 4.8 & 41.1 $\pm$ 4.9& 8.51 & 8.66 $\pm$ 0.16 & 0.76 $\pm$ 0.10 & 0.43 $\pm$ 0.06 \\	
		\hline
		\hline
	\end{tabular}
	\caption{As in Table \ref{table_sn}, but for other actinides}
	\label{table_sn2}
\end{table*}

\end{document}


\title{Supplemental Material: \\ Nuclear state and level densities of actinides in the shell-model Monte Carlo}
	
	\author{D. DeMartini and Y. Alhassid}
	
	\affiliation{Center for Theoretical Physics, Sloane Physics Laboratory, Yale University, New Haven, Connecticut 06520, USA}
	

	\maketitle

\subsection{Ground-state energies of odd-mass nuclei}
	
	The SMMC has a Monte-Carlo sign problem at low temperatures when projecting onto an odd number of protons and/or neutrons, even for good-sign interactions. While the Monte Carlo sign remains near unity for moderate temperatures near the neutron resonance energies, it quickly falls as the temperature decreases. Typically, the sign $\langle \Phi \rangle \approx 1$ for $\beta \le 3$ MeV$^{-1}$,  but it rapidly decreases at larger values of $\beta$ and uncertainties of most observables become too large to be useful around $\beta \sim 7-8$ MeV$^{-1}$. For odd-mass actinides, which typically have several excited states below $\sim 200$ keV, this makes it impossible to calculate directly their ground-state properties within the SMMC.
	
	A precise determination of the ground-state energy is necessary in order to compute NSDs that are functions of the excitation energy $E_x = E - E_0$. In Ref.~\cite{Ozen2015}, this difficulty was overcome by augmenting the SMMC calculations with experimental data on low-lying excited states of the odd-mass nuclei of interest. While this method works well, one would like to have a self-contained method that does not rely upon experimental data. To this end, the partition function extrapolation method (PFEM) was developed in Ref.~\cite{Alhassid2024}. 
	
	The PFEM determines the ground-state energy of a system given its thermal energy $E(\beta)$ at moderate temperatures and an appropriate parameterized model of its state density. The PFEM starts by determining the excitation partition function $Z'$ of the system relative to a chosen reference energy $E_{\textrm{ref}}$
	\begin{equation}
		Z'(\beta;E_{\textrm{ref}}) = Z(\beta)e^{E_{\textrm{ref}}\beta},
	\end{equation}
	where $Z(\beta) = \textrm{Tr} e^{-\beta \hat{H}}$ is the partition function in terms of absolute energies. This excitation partition function is related to the excitation partition function relative to the ground-state energy $E_0$ by
	\begin{equation} \label{eq:lnZ}
		\ln Z'(\beta;E_{\textrm{ref}}) = \ln Z'(\beta;E_0) - \beta(E_0 - E_{\textrm{ref}}).
	\end{equation}
	
	The excitation partition function with respect to $E_0$ can be related to a parameterized model of the state density $\rho(E_x)$ by the Laplace transform
	\begin{equation} \label{eq:Z}
		Z'(\beta;E_0) = \int_0^{\infty} dE_x \rho(E_x) e^{-\beta E_x}.
	\end{equation}
	
	\begin{figure}
		\includegraphics[width=\linewidth]{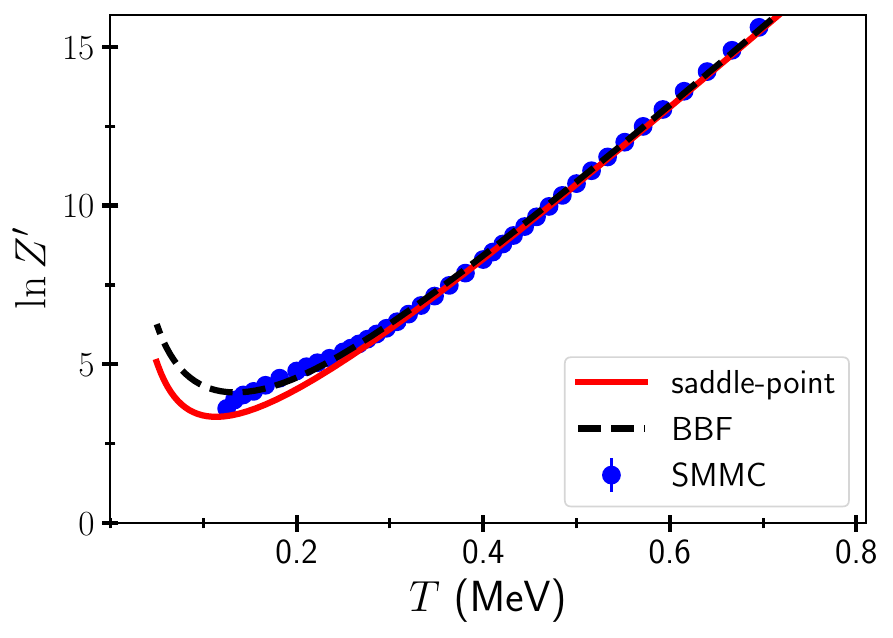}
		\caption{The logarithm of the excitation partition function $\ln Z'$ of $^{239}$U with $E_{\textrm{ref}} = -183$ MeV as a function of the excitation energy $E_x$.  The SMMC results (blue circles) are compared with the saddle-point fit (red line) and the BBF fit (black dashed line).}
		\label{fig_lnZ}
	\end{figure}
	
	The parameterized model we use is the back-shifted Bethe formula (BBF). Inserting the BBF into Eq.~(\ref{eq:Z}) and then using  the corresponding $Z'(\beta,E_0)$ in  Eq.~(\ref{eq:lnZ}) enables us to perform a $\chi^2$ fit to determine $E_0$ as well as the two parameters of the BBF, $a$ and $\Delta$. In practice, the fit is carried out in two steps. In the first step a saddle-point approximation is applied to Eq.~(\ref{eq:Z}) with the BBF for $\rho(E_x)$ to give
	\begin{equation}
		\ln Z'(\beta;E_{\textrm{ref}}) \approx \frac{a}{\beta} + \ln \left( \frac{\pi \beta}{6a} \right) - \beta S,
	\end{equation}
	where $S = E_0 - E_{\textrm{ref}} + \Delta$.  The SMMC data is used to calculate $\ln Z'(\beta;E_{\textrm{ref}})$ at moderate temperatures where the Monte-Carlo sign is still good and the BBF is expected to be a good model of the data. Then, a two-parameter fit is used to determine $a$ and $S$. In the second step, with $a$ and $S$ held constant, the integral in Eq.~(\ref{eq:Z}) is carried out numerically using the full BBF and a one-parameter fit to the lowest temperature SMMC data available is performed to determine the value of $E_0$ (the value of $\Delta$ is then determined by $\Delta=S-(E_0-E_{\rm ref})$). Figure \ref{fig_lnZ} shows an example of these two fits for $^{239}$U. 
	
	The PFEM was tested in even-mass lanthanides (against direct SMMC calculations of $E_0$) and used successfully in odd-mass lanthanides for which it  produced ground-state energies consistent with those found using experimental data~\cite{Alhassid2024}. Here we have performed the PFEM on the odd-mass actinides using SMMC data at inverse temperatures up to $\beta = 8$ MeV$^{-1}$. Table \ref{tab_oddgs} summarizes the ground-state energies of both the even-even and odd-mass actinides. The PFEM is able to determine the ground-state energies with uncertainties that are a few times as large as the direct fits of the thermal energy in the even-even nuclei. We note that the low-temperature observables in the odd-mass actinides were calculated using four times as many Monte-Carlo samples ($32000$) as compared to the even-even nuclei ($8000$).  
	
	\begin{table}
		\begin{tabular}{c|c}
			\hline
			Nucleus & $E_0$ (MeV)  \\
			\hline
			$^{232}$Th & $-135.17 \pm 0.07$ \\
			$^{234}$U & $-147.57 \pm 0.03$ \\
			$^{235}$U & $-154.77^{+0.12}_{-0.09}$ \\
			$^{236}$U & $-162.52 \pm 0.03$ \\
			$^{237}$U &    $-169.31^{+0.17}_{-0.18}$  \\
			$^{238}$U & $-176.82 \pm 0.05$ \\
			$^{239}$U &    $-183.22^{+0.10}_{-0.07}$  \\
			$^{240}$Pu & $-189.71 \pm 0.03$ \\
			$^{241}$Pu &   $-197.33^{+0.09}_{-0.02}$ \\
			$^{242}$Pu & $-205.63 \pm 0.06$ \\
			$^{243}$Pu &   $-212.86^{+0.16}_{-0.34}$ \\
			$^{246}$Cm & $-234.64 \pm 0.05$ \\
			$^{247}$Cm & $-242.39^{+0.18}_{-0.04} $ \\
			$^{248}$Cm & $-251.07 \pm 0.06$ \\
			$^{250}$Cf & $-263.50 \pm 0.05$ \\
			\hline
		\end{tabular}
		\caption{Ground-state energies $E_0$ of selected actinides calculated with the SMMC. Values of even-even nuclei were determined from direct SMMC calculations of the thermal energy $E(\beta)$ at low temperatures, while those for odd-mass nuclei were determined using the PFEM.}
		\label{tab_oddgs}
	\end{table}

	\subsection{Experimental level densities at $S_n$}
	
Table I of the main text compares the NLDs at the neutron separation energies $\tilde{\rho}(S_n)$ computed with the SMMC to those presented in the works by the Oslo group~\cite{Guttormsen2013,Laplace2016,Zeiser2019}. While these values are presented as ``experimental'' in the table, they have been calculated using a combination of experimentally measured average neutron resonance spacings $D_0$ together with empirical models of the spin distribution and, in particular, the spin-cutoff model.
In the following we discuss the method used by the Oslo group to calculate these ``experimental" values of $\tilde \rho(S_n)$.
	
The NLDs at $S_n$ are calculated using Eq.~(12) of the main text, an equation derived from the spin-cutoff model. The values of $D_0$ and $J_t$ are taken from experiments, while the spin-cutoff parameter $\sigma(E_x)$ is given by the following empirical formula based on a rigid-body moment of inertia~\cite{vonEgidy2005}
	\begin{equation}
		\sigma^2(E_x) = 0.0146A^{5/3}\frac{1+\sqrt{4a(E_x-\Delta)}}{2a},
		\label{eq_spin}
	\end{equation}
where $a$ is the level density parameter and $\Delta$ is the backshift parameter of the BBF. 

For the nuclei in Table II of the main text, where Oslo method experiments have not been performed, we use the method of Ref.~\cite{Ozen2025} to determine the spin-cutoff parameter, i.e., 
\begin{equation}
	\sigma^2(E_x) = \frac{I T_{\text{eff}}}{\hbar^2}, \,\,\,\,\, T_{\text{eff}} = \sqrt{\frac{E_x - \Delta}{a}},
	\label{eq_spin2}
\end{equation}
where at the neutron separation energies $S_n$, it is assumed that the nucleus has a rigid-body moment of inertia given by $I_{\text{rig}}/\hbar^2 = \frac{2}{5} m r_0^2 A^{5/3}$ MeV$^{-1}$, where $r_0 \approx 1.25$ fm is the nuclear radius parameter and $m=939$ MeV is the nucleon mass. We use this formula for the nuclei in Table II of the main text for consistency; the two formulas give values of $\sigma(S_n)$ that differ by less than a fraction of a percent in all cases. For all nuclei a 10\% uncertainty in $\sigma$ has been assumed when propagating the uncertainty to the level density. 

The experimental values of $\tilde{\rho}(S_n)$ shown in Table II of the main text have been calculated using the $\sigma$ values determined by Eq.~(\ref{eq_spin}) with the BBF parameters from Ref.~\cite{Ozen2025} for nuclei with no Oslo data.  For the nuclei where Oslo method results are available, the values of $\tilde{\rho}(S_n)$ are taken from their respective works~\cite{Guttormsen2013,Laplace2016,Zeiser2019}.  We note that the authors of these experimental results use the same method described here but with different values of the BBF parameters than those cited in Ref.~\cite{Ozen2025}.

It was pointed out in Ref.~\cite{Larsen2011} that different models of the spin-cutoff parameter can lead to significant variations in the predicted value of $\tilde{\rho}(S_n)$. The models of Refs.~\cite{vonEgidy2005} and \cite{Gilbert1965} give differences of up to $\sim 50\%$, while the HFB-plus-combinatorial model~\cite{Goriely2008} typically differs by a factor of 2 to 3 compared to those models. 
	
	\subsection{Hartree-Fock-Bogoliubov approximation and pairing gaps}

The relative computational simplicity of mean-field approximations makes them particularly useful for computing quantities in broad regions of the nuclear chart. 
These techniques remain useful despite the fact they often miss important correlations (as can be seen for example in comparing the NSDs computed in the mean-field approximation versus the NSDs calculated in SMMC). In this work, a finite-temperature Hartree-Fock-Bogoliubov (HFB) approximation~\cite{Goodman1981} was utilized for testing efficiently interaction parameters of the Hamiltonian and to serve as a point of comparison with the SMMC. The main details of the HFB approximation are outlined below. For a more complete discussion of the HFB approximation and the HF-SHELL code we use, see in Ref.~\cite{Ryssens2021}.
	
	\subsubsection{The HFB approximation}
	
	The HFB approximates the density matrix of a general Hamiltonian containing  a two-body interaction with a density matrix of the form
	\begin{equation}
		\hat{D}_{\textrm{HFB}} = \frac{1}{Z_{\textrm{HFB}}} \prod_{\nu=p,n} e^{-\beta \hat{K}_{\nu} + \alpha_{\nu} \hat{N}_{\nu}},
	\end{equation}
	where $\hat{K}_{\nu}$ are one-body operators with the bilinear form
	\begin{gather}
		\hat{K}_{\nu} = \frac{1}{2} \begin{pmatrix}
			\hat{a}_{\nu} \\
			\hat{a}_{\nu}^{\dagger}
		\end{pmatrix} ^{\dagger} \mathcal{K}_{\nu}  \begin{pmatrix}
		\hat{a}_{\nu} \\
		\hat{a}_{\nu}^{\dagger}
	\end{pmatrix} \;.
	\end{gather}
	Here $\mathcal{K}_{\nu}$ is the $2M_{\nu} \times 2M_{\nu}$ matrix representation of the operator $\hat{K}_{\nu}$, $M_{\nu}$ is the number of single-particle states of nucleon species $\nu$ ($\nu =p$ or $n$), and $\alpha_{\nu} = \beta \mu_{\nu}$. The corresponding HFB partition function is $Z_{\textrm{HFB}} = \prod_{\nu=p,n} \textrm{Tr}\, e^{-\beta \hat{K}_{\nu} + \alpha_{\nu} \hat{N}_{\nu}}$. 
	
	The one-body density matrix $\rho_{\nu}$ and anomalous density matrix $\kappa_{\nu}$ are defined by
	\begin{equation}
		\rho_{\nu,ij} = \textrm{Tr} \, \left( \hat{D}_{\textrm{HFB}} \hat{a}_j^{\dagger} \hat{a}_i \right), \,\,\,\, \kappa_{\nu,ij} = \textrm{Tr} \left( \hat{D}_{\textrm{HFB}} \hat{a}_j \hat{a}_i \right). 
	\end{equation}
	These matrices are determined by minimization of the grand-canonical potential at given temperature $T$ and chemical potentials $\mu_p$ and $\mu_n$
	\begin{equation}
		\Omega(T,\mu_p,\mu_n) = E -TS -\mu_p N_p - \mu_n N_n,
	\end{equation}
	where the average energy $E$, entropy $S$, and average particle numbers $N_{\nu}$ are given by
	\begin{equation}
		\begin{aligned}
			E &= \textrm{Tr}\left(\hat{D}_{\rm HFB} \hat{H}\right), \\
			S &= - \textrm{Tr}\left(\hat{D}_{\rm HFB} \ln \hat{D}_{\rm HFB}\right), \\
			N_{\nu} &= \textrm{Tr}\left(\hat{D}_{\rm HFB} \hat{N}_{\nu}\right). \\
		\end{aligned}
	\end{equation}
The chemical potentials are determined to reproduce the average particle number of each nucleon species for the given nucleus. With the energy $E$ and entropy $S$ determined as functions of the temperature, the NSD in the HFB approximation can be calculated with a similar saddle-point formula as in the SMMC [see Eq.~(7) in the main text]. As was shown in the main text, the HFB NSD of deformed nuclei is significantly lower than the SMMC NSD. The HFB only reproduces the intrinsic states and not the rotational bands built on top of them in deformed nuclei.
	
	\subsubsection{Pairing gaps and mass staggerings}
	
	The three-point odd-even staggering (OES) formula for a nucleus with with $Z$ protons and $N$ neutrons is 
	\begin{equation}
		\textrm{OES}_n = \frac{(-1)^N}{2}(B(Z,N-1) + B(Z,N+1) - 2B(Z,N)),
	\end{equation} 
	where $B(Z,N)$ is the (negative) binding energy of a nucleus. An analogous expression defines the odd-even staggering $\textrm{OES}_p$ along an isotonic chain of differing proton number $Z$~\cite{Krieger1990,Cwiok1996}. 
	
	Within the HFB, the pairing gap matrix elements $\Delta_{\nu,ij}$ are defined by~\cite{Goodman1981}
	\begin{equation}
		\Delta_{\nu,ij} = \frac{1}{2} \sum_{kl=1}^{M_{\nu}} \bar{v}^{\nu \nu}_{ijkl} \kappa_{\nu,kl},
	\end{equation}
	where $\bar{v}^{\nu \nu}_{ijkl}$ are the angular-momentum uncoupled two-body matrix elements between nucleons of the same species. The pairing gap matrix elements are orbital dependent and there is no unique definition of the pairing gap for general interactions. Two different forms of the average pairing gap are computed by HF-SHELL~\cite{Ryssens2021}
	\begin{equation}
		\begin{aligned}
			\langle v^2 \Delta \rangle_{\nu} &= \frac{\sum_{ij=1}^{M_{\nu}}\rho_{\nu,ij}\Delta_{\nu,ji}}{\textrm{tr}_{\nu} \rho_{\nu}},\\
			\langle uv \Delta \rangle_{\nu} &= \frac{\sum_{ij=1}^{M_{\nu}}\kappa_{\nu,ij}\Delta_{\nu,ji}}{\sum_{ij=1}^{M_{\nu}} |\kappa_{q,ij}|},
		\end{aligned}
	\end{equation}
	where $\textrm{tr}_{\nu}$ indicates a trace over only the single-particle space of nucleon species $\nu$. The two different definitions of the average pairing gap amount to two different choices of weights for the single-particle orbitals. In this work $\langle uv \Delta \rangle_{\nu}$ was used, but in practice the two definitions typically give results that differ by only a few percent.  
	
	Mean-field approximations miss important correlations and care must be taken when comparing their results to experiments. In Ref.~\cite{Donati2005} it was found that, in deformed nuclei, the mean-field approximation on its own accounts for only about half of the experimentally observed OESs. We therefore estimate the OES in the HFB by 
	\begin{equation}\label{OES}
		\textrm{OES}_{\nu}^{(\textrm{HFB})} = 2 \langle uv\Delta \rangle_{\nu}.
	\end{equation}

\begin{figure*}
		\includegraphics[width=\linewidth]{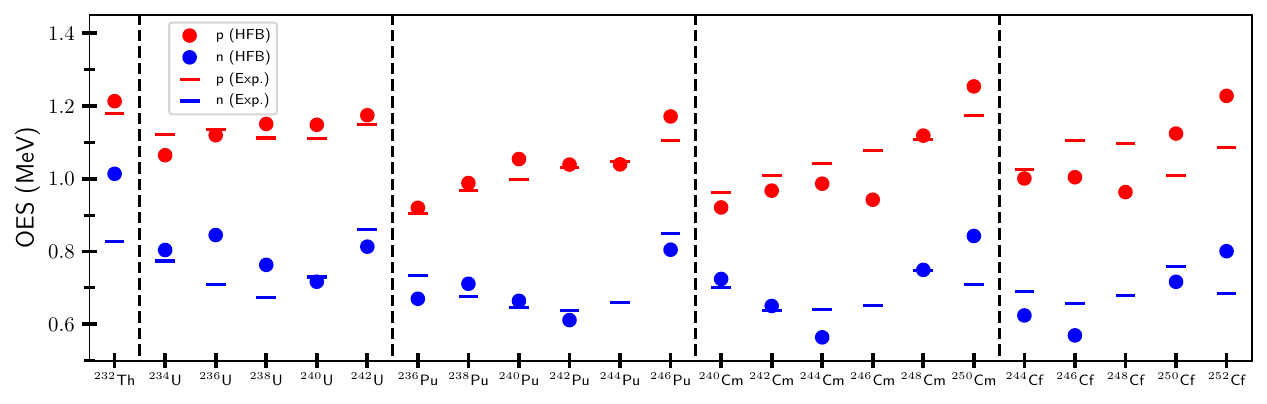}
		\caption{Comparison between the OESs from experiments (bars) and those estimated in the HFB (circles) for several chains of actinide isotopes.}
		\label{fig_OES}
	\end{figure*}

	Comparing these pairing gap calculations to the experimental OES data~\cite{Capote2009} was an important consideration in determining the interaction parameters for the actinides. Various functional forms of the interaction parameters $g_p$, $g_n$, and $k_2$ with a few free parameters each were tuned by fitting the OESs calculated in the HFB to the experimental values across a range of nuclei. The pairing gaps for the final interaction parameters are shown in Fig.~\ref{fig_OES}, where the theoretical values (\ref{OES}) (circles) are compared with the experimental values (bars) for both $\nu=p,n$.

	A few nuclei ($^{244}$Pu, $^{246}$Cm, and $^{248}$Cf) have anomalous, near-zero values of the neutron paring gaps in the HFB. These nuclei were not included in the fits used to determine the interaction parameters. The average magnitude of the difference between the experimental and HFB staggerings $|\textrm{OES}_{\nu}^{\textrm{HFB}} - \textrm{OES}_{\nu}^{\textrm{exp.}}|$ is 55 keV for protons and 73 keV for neutrons, excluding the anomalous nuclei for the neutrons.  
	
	The good agreement between the theoretical and experimental staggerings, along with the good agreement between the NLDs for the few nuclei where experimental data is available suggest that the model space and interaction parameters discussed here can accurately reproduce the NSDs and NLDs for all of the nuclei in Fig.~\ref{fig_OES}. 
	
\subsection{Moment of Inertia and the Spin-Cutoff Parameter}

\begin{figure}
	\includegraphics[width=\linewidth]{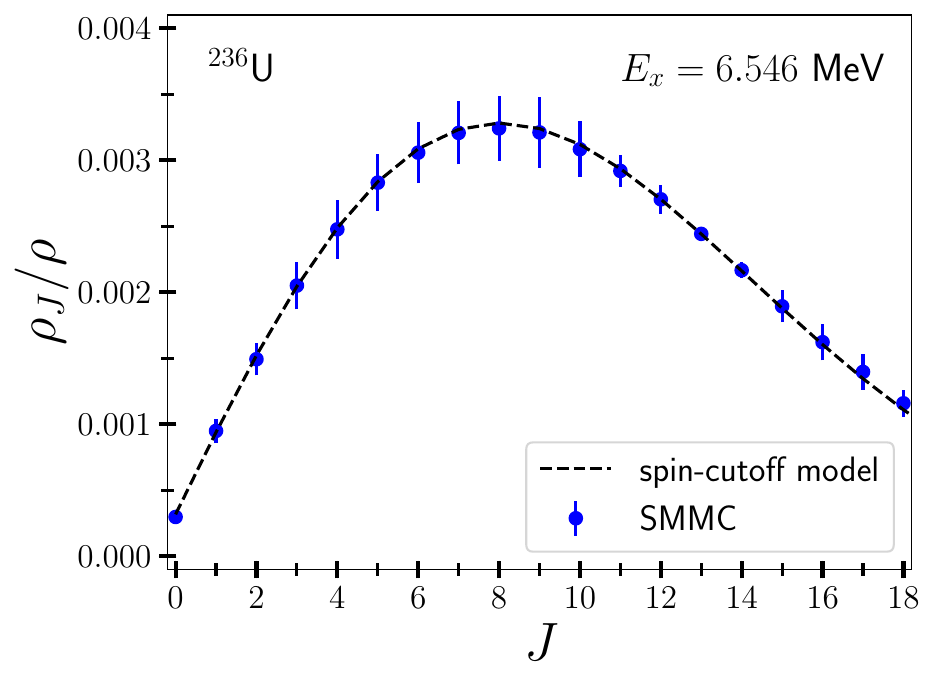}
	\caption{The spin distribution $\rho_J/\rho$ in $^{236}$U at its neutron separation energy $E_x=S_n=6.546$ MeV calculated with the SMMC (solid circles with error bars) and its fit to the spin-cutoff model (dashed line).}
	\label{fig_spin}
	\end{figure}

	The spin-cutoff parameter at the neutron separation energy $\sigma(S_n)$ is a necessary quantity for determining NLDs experimentally as it directly relates the NLDs at that energy to the measured average s-wave neutron resonance spacing $D_0$. In the SMMC, the spin-cutoff parameter is determined by fitting the calculated spin distribution to the  distribution in the spin-cutoff model [Eq.~(10) in the main text]. Figure \ref{fig_spin} shows such a fit (dashed line) to the SMMC data (circles with error bars) for $^{236}$U. All of the nuclei studied here show similarly excellent fits to the spin-cutoff model at $S_n$.  
	
	\begin{figure}
	\includegraphics[width=0.9\linewidth]{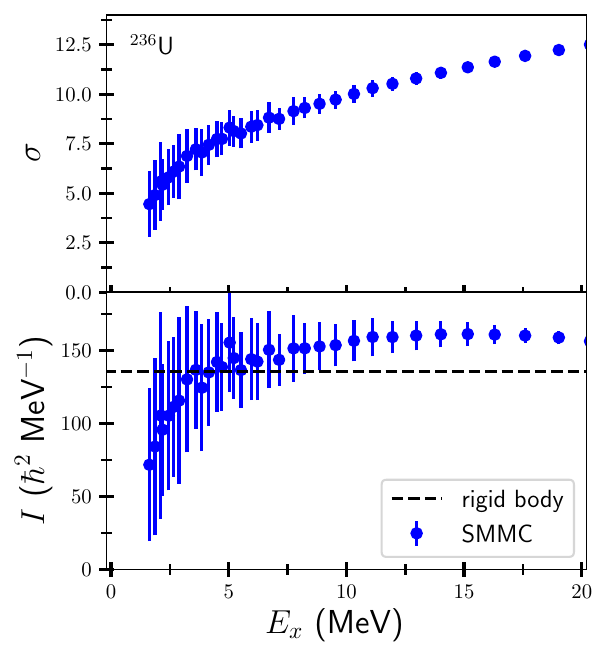}
	\caption{Spin-cutoff parameter $\sigma$ calculated from Eq.~(\ref{eq_sig}) (top panel) and thermal moment of inertia $I$ (bottom panel) of $^{236}$U as functions of the excitation energy $E_x$. The dashed line is the rigid-body moment of inertia $I_{\text{rig}}$.}
	\label{fig_moi}
\end{figure}

	The energy dependence of $\sigma$ also provides useful information on pairing correlations. While $\sigma(E_x)$ can be calculated by fits to the spin distribution at a given excitation energy $E_x$, it can also be calculated from the ratio of the NLD and NSD
	\begin{equation}
		\frac{\tilde{\rho}(E_x)}{\rho(E_x)} = \frac{1}{\sqrt{2\pi}\sigma(E_x)}.
		\label{eq_sig}
	\end{equation}
	
	A  thermal moment of inertia $I$ can be calculated from $\sigma$ using the relation
	\begin{equation}
		I = \frac{\sigma^2 \hbar^2}{T},
		\label{eq_I}
	\end{equation}
	where $T$ is the temperature of the nucleus such that it has an average excitation energy $E_x = E - E_0$. At higher energies, when pairing effects are weak, the moment of inertia is expected to be close to its rigid-body value $I_{\text{rig}}$ In Fig.~\ref{fig_moi}, we show the calculated spin-cutoff parameter (top panel) and moment of inertia (bottom panel) of $^{236}$U as a function of the excitation energy $E_x$ using Eqs.~(\ref{eq_sig}) and (\ref{eq_I}). 
	
 At higher excitation energies, the moment of inertia is approximately constant and slightly above the rigid-body value, and it decreases rapidly at low energies. This decrease  has only been observed in even-even nuclei~\cite{Alhassid2007} and is attributed to the onset of strong pairing correlations at low excitation energies. 
	
\bibliography{actinideNLDs}